\documentstyle[aaspp4,12pt,epsfig,color]{article}
\begin{document}

\title{Cloning Hubble Deep Fields I: A Model-Independent Measurement\\
	 of Galaxy Evolution}

\author{Rychard Bouwens} 
\affil{Physics Department,
   University of California,
    Berkeley, CA 94720; bouwens@astro.berkeley.edu}
\author{Tom Broadhurst}
\affil{Astronomy Department,
    University of California,
    Berkeley, CA 94720; tjb@astro.berkeley.edu}
\centerline \&
\author {Joseph Silk}
\affil{Astronomy and Physics Departments, and Center for Particle
    Astrophysics, University of California,
     Berkeley, CA 94720; silk@astro.berkeley.edu}

\begin{abstract}

We present a model-independent method of quantifying galaxy evolution
in high-resolution images, which we apply to the Hubble Deep Field
(HDF).  Our procedure is to k-correct all pixels belonging to the
images of a complete set of bright galaxies and then to replicate each
galaxy image to higher redshift by the product of its space density,
$1/V_{max}$, and the cosmological volume.  The set of bright galaxies
is itself selected from the HDF, because presently the HDF provides
the highest quality UV images of a redshift-complete sample of
galaxies (31 galaxies with $I<21.9$, $\bar{z}=0.5$, and for which
$V/V_{max}$ is spread fairly). These galaxies are bright enough to
permit accurate pixel-by-pixel k-corrections into the restframe UV
($\sim 2000$ $\AA$). We match the shot noise, spatial sampling and PSF
smoothing of the HDF data, resulting in entirely empirical and
parameter-free ``no-evolution'' deep fields of galaxies for direct
comparison with the HDF.  In addition, the overcounting rate and the
level of incompleteness can be accurately quantified by this
procedure.  We obtain the following results.  Faint HDF galaxies
($I>24$) are much smaller, more numerous, and less regular than our
``no-evolution'' extrapolation, for any interesting geometry.  A
higher proportion of HDF galaxies ``dropout'' in both $U$ and $B$,
indicating that some galaxies were brighter at higher redshifts than
our ``cloned'' $z\sim0.5$ population.
 
\end{abstract}

\keywords{galaxies: evolution --- galaxies: scale-lengths}

\section{Introduction}

 Amongst the highest quality data available for exploring galaxy
evolution are the long exposures of the Hubble Deep Field (Williams et
al.\ 1996), which register the faintest and sharpest images of
galaxies ever detected.  The counts of faint galaxies are found to
increase to the completeness limit ($I \sim 28$), roughly doubling in
number per magnitude, $d\log(N)/dm=0.2-0.3$ (Williams et al.\ 1996;
see Tables 9-10), in all bands.  This trend is accompanied by a marked
decline in angular size, so that most faint galaxies are barely
resolved.  While much of this light must arise from the epoch when
galaxies and stars were first forming, the photometric limit of the
HDF far exceeds the practical limit of spectroscopy (some $97\%$ of
the HDF galaxies do not have measured redshifts), hence for the
foreseeable future interpretation of these images rests principally on
photometric information.

A major impediment in analysing the HDF images is our ignorance of the
ultraviolet (UV) properties of ordinary galaxies, since at high
redshift it is the restframe UV light that is detected in the optical
bands. Despite the rather modest UV performance of imaging satellites
it has become apparent that the appearance of even ordinary
Hubble-sequence galaxies is knotty and irregular in the UV, governed
by the spatial distribution of high contrast star-forming regions
(O'Connell \& Markum 1996; Giavalisco et al.\ 1996a).  Hence, it is
unclear to what extent the large numbers of blue and irregular looking
objects reported in HST images (Glazebrook et al.\ 1995; Driver,
Windhorst, \& Griffiths 1995; Abraham et al.\ 1996b; Colley et al.\
1997) simply result from redshifted UV light rather than evolution.

 There have been several efforts to obtain images of representative
nearby galaxies. The FOCA balloon experiment at 2000 $\AA$ has
measured six UV images of local galaxies (Blecha et al.\ 1990), the
astro-1 (astro-2) mission measured 27 (45) UV images of objects using
the Ultraviolet Imaging Telescope (Stecher et al.\ 1992), and Maoz et
al.\ (1997) have recently compiled UV images of 110 galaxies using the
FOC on HST ($2300 \AA$).  Cognizant of the importance of these UV
images for assessing galaxy evolution, researchers (Bohlin et al.\
1991; Giavalisco et al.\ 1996a; Maoz 1997) have already used these
data to comment on deep HST images.  While these studies have been
very illustrative, the samples are either incomplete or register only
the inner regions of galaxies so that a quantification of galaxy
evolution in deep HST images has hitherto not been attempted.

 Due to the inadequacies of UV instrumentation, it is interesting to
note that even the gross effect of the Lyman-limit is more easily observed
in the {\it optical}, by virtue of the high redshifts of some
faint galaxies (Steidel et al.\ 1996a). The restframe UV morphologies
of several tens of these high redshift galaxies are measured directly with HST,
particularly in the Hubble Deep Field (Steidel et al.\ 1996b;
Giavalisco et al.\ 1996b; Lowenthal et al.\ 1997) down to the Lyman
limit.  These galaxies lie in the range $z\sim2.0-3.5$ and are
reported to have a generally compact appearance, but many have faint
plumes and some are up to 3'' in length (object C4-06 of Steidel et
al.\ 1996b).  The UV continuum flux places some constraint on the
luminosity density of star-forming galaxies at high-z (Madau et al.\
1996) at least for galaxies blue enough to be dominated by the light
of O-stars. Comparison of the implied high mass star-formation rate
with local estimates from from H$\alpha$ emission (Gallego et al.\
1995), provides an indirect means of estimating the evolution of the
integrated UV-luminosity density (Madau et al.\ 1996).

In this paper, we outline a simple and practical method for measuring
evolution in deep space images, by constructing empirical
``no-evolution'' fields of galaxies as a benchmark, matched in image
properties to the data, in this case the HDF.  Our method has the
distinct advantages of being purely empirical and entirely free of
parameterizations, so that our conclusions are liberated from the
usual caveats.  We first establish the k-correction for each pixel
belonging to a given galaxy image drawn from a complete sample of
relatively bright galaxies with redshifts (selected from the HDF).  We
then scale the numbers of these k-corrected galaxy images out to high
redshift in proportion to the product of the volume and the space
density of each galaxy, the latter being set by its value of
$1/V_{max}$.  Finally, we correct the PSF, add the appropriate noise
and sampling of the HDF observations and compare these
``no-evolution'' fields with the HDF at fainter magnitudes ($I>24$).
This approach is feasible due to the relatively high quality $F300W$
images of the brighter galaxies detected in the HDF ($I<22$), as well
as the many subsequent redshift measurements of HDF galaxies with the
Keck Telescope (Cohen et al.\ 1996; Lowenthal et al.\ 1997).  Together
these observations allow us to construct a statistically useful
complete sample of 31 galaxies with sufficient UV-optical coverage for
construction of an entirely empirical and parameter free
``no-evolution'' deep field for comparison with the HDF. The universal
geometry is the only unknown, dictating the volume and angle-redshift
relations, and bearing somewhat on the degree of evolution derived in
the $I$ band.
 
By following this procedure, we make no recourse to the usual
parameterizations of morphological or colour classes, galaxy
luminosity functions, and light profiles.  Implicitly, such
parameterizations have tended to assume one-parameter relationships,
particularly between morphological type and spectral type, reducing
the inherent richness of the general galaxy population and potentially
introducing artificial model dependencies. In light of this, it is not
very surprising that simulations of two-dimensional images look
artificial, appearing more ``hygienic'' than reality, and hence of
limited value in reliably quantifying evolution.  Designating a
morphological class has in itself proven to be problematic for
well-resolved galaxies.  Naim et al.\ (1995), for example, have found
from a comparison of the classification of six human `experts' that
classification exhibits a variance of 2 morphological types relative
to one another.  Hence there is still no clear way to map an observed
image onto a given morphological type or to produce a realistic image
given a morphological type.

Finally, we emphasize that our procedure is free of the many internal
selection biases which have long haunted low-redshift to high-redshift
comparisons, e.g., where deep CCD data have lower surface brightness
thresholds than photographic surveys used to construct the local
luminosity functions.

We begin this paper by describing the manner in which we select our
bright galaxy sample in \S2.  In \S3, we discuss our empirical
procedure for deriving pixel-by-pixel spectral energy distributions
and number densities for each of these bright galaxies.  In \S4, we
assess both the fairness of our sample and the viability of the method
by making some simple predictions chiefly with regard to the Canada
France Redshift Survey (CFRS; Lilly et al.\ 1995).  In \S5, we
describe the simulation procedure in detail, and in \S6 we describe
our analysis of the simulations and the HDF data using SExtractor.  In
\S7 we present our results, in \S8 we discuss these results, and in
\S9 we present a summary.

We adopt $H_0 = 50\,\textrm{km/s/Mpc}$ and express all magnitudes in
this paper in the AB\footnote{$m \textrm{(AB)} = -2.5 \log f_{\nu}
\textrm{(erg/cm/cm/s/Hz)} - 48.60$ (Oke 1974)} magnitude system
(defined in terms of a flat spectrum in frequency).  Also, to
associate the HDF bands with their more familiar optical counterparts,
we shall refer to the $F814W$, $F606W$, $F450W$, and $F300W$ bands as
$I_{814}$, $V_{606}$, $B_{450}$, and $U_{300}$, respectively,
throughout this paper.

\section{Bright Galaxy Sample}

We select our sample of bright galaxies from the HDF. First we measure
the photometry of all galaxies detected in the HDF by applying the
SExtractor photometry package 1.2b5 (Bertin \& Arnouts 1996) to the
publicly released version 2 images of the HDF.  In using SExtractor we
require objects to be at least 2 $\sigma$ above the sky noise in the
$I_{814}$ image over an area of at least 10 contiguous pixels, after
first mildly smoothing the images by a Gaussian of width 0.06-arcsec
sigma (the approximate width of the PSF).  We take the deblending
parameter ($DEBLEND\_MINCONT$) to be 0.04, which we found to be a good
compromise between splitting too many faint objects apart and merging
too many different objects together.  For photometry, we use isophotal
magnitudes ($MAG\_BEST$) corrected with a Gaussian extrapolation to
approximate the light beyond the isophotes for all non-crowded (almost
all) objects, where we take our isophotes to be equal to 2 $\sigma$
times the sky noise ($\sim24.9\,\textrm{mag/arcsec}^2$).  Objects with
half-light radii less than 0.15 arcsec are excluded as stars (5
objects with $I_{814,AB} < 22.33$). Only objects on the three Wide
Field Camera chips lying within 4 arcsecs of the edges are included,
beyond which the image quality is diminished.

We match up our resulting photometric catalogues redshifts measured by
various groups (Cohen et al.\ 1996; Lowenthal et al.\ 1997) to
construct a strictly magnitude-limited subset, bright enough to be
highly complete in redshift.  We settle for a conservative limit of
$I_{814,AB} < 22.33$, which contains 32 objects in total -- only one
of which is without a redshift and for which we know of no attempt to
measure a redshift nor any reason to suppose that its measurement
would prove difficult.  Consequently, we shall assume that the 31
objects with redshifts are representative, and simply scale volume
densities by $32/31$. As more redshifts become available this analysis
can be extended, although clearly the limited area of the HDF means
that new fields are required in order to significantly enlarge the
bright galaxy sample and to help average over the possible effects of
clustering.

The resultant bright sample is listed in Table 1, along with our
determination of its coordinates, $I_{814}$ magnitude, absolute
magnitude, surface brightness (taken to be $m_{b_J ^0}/(2 \pi r_{hl}
^2)$), (where $m_{b_J}$ is the rest-frame $b_J$ apparent magnitude and
$r_{hl}$ is the half-light radius), our eye-ball determination of the
morphological type, and the known redshift.  For reference we also
include our determination of the k-correction over the surface of each
galaxy at $z=2.5$.  The calculation of k-corrections, rest-frame
absolute magnitudes, and rest-frame surface brightnesses are described
below.  In addition, the redshift distribution of our bright subset is
shown in Figure 1, for comparison with all of the redshifts measured
in the compilation by Cohen et al.\ (1996), which includes the HDF and
its flanking fields.  We display the images of the bright sample in
Figures 2a and Figure 2b, as a set of colour images ($B_{450}$,
$V_{606}$, and $I_{814}$ bands) as they appear if k-corrected to the
mean bright sample redshift, $z=0.5$, and also to $z=2.5$, using the
pixel-by-pixel k-corrections described in the next section.

\begin{figure}
\epsscale{0.95}
\plotone{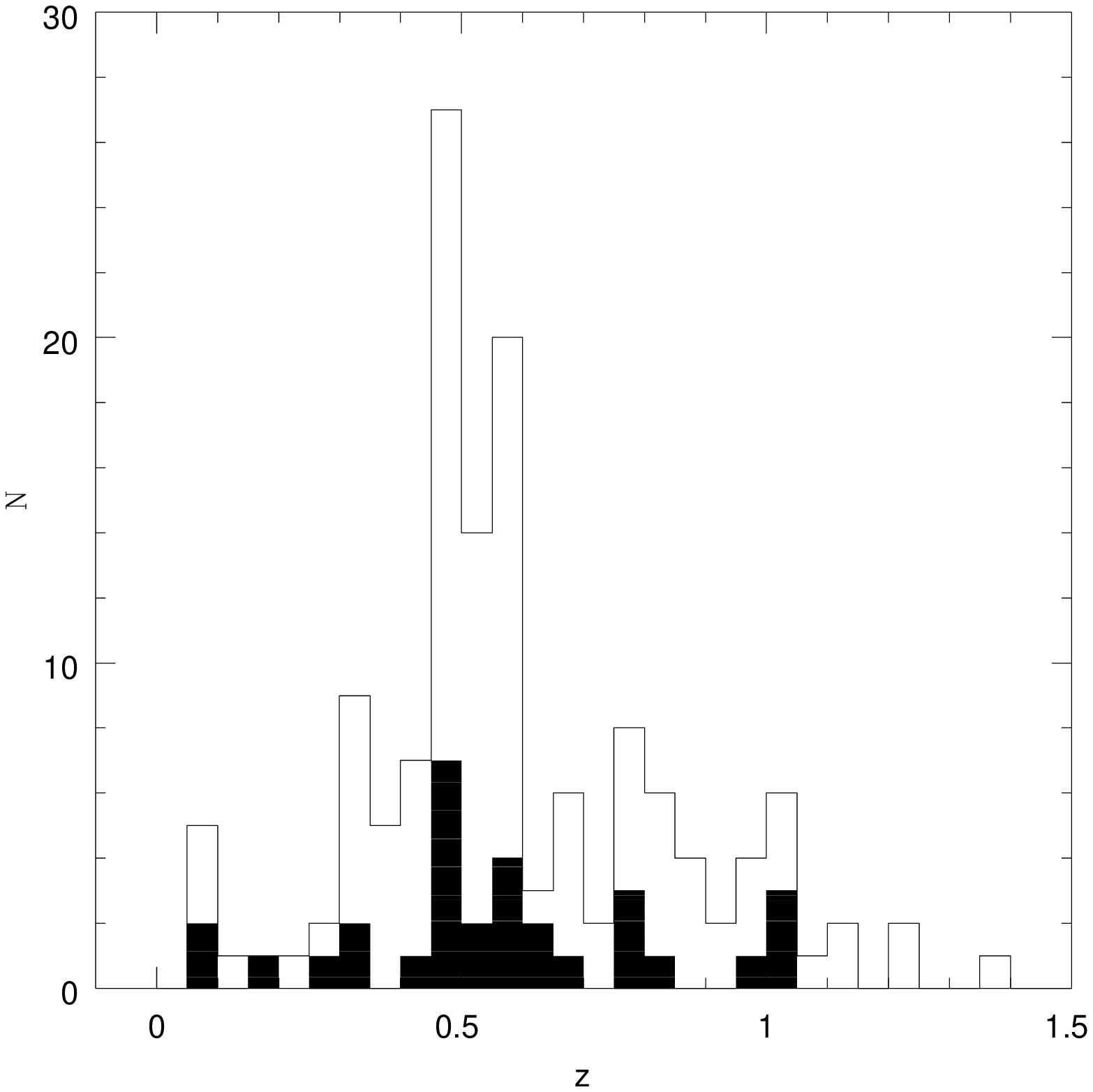}
\caption{Redshift distribution of our magnitude-limited sample
($I_{814,AB} < 22.33$) from the HDF follow-up redshift surveys (solid
histogram) plotted along with the redshift distribution of all the
redshifts available to this limit compiled by Cohen et al.\ (1996) of
the HDF and flanking fields.\label{samplez}}
\end{figure}

\begin{figure}
\begin{center}
\resizebox{12cm}{!}{\includegraphics*[25,50][660,760]{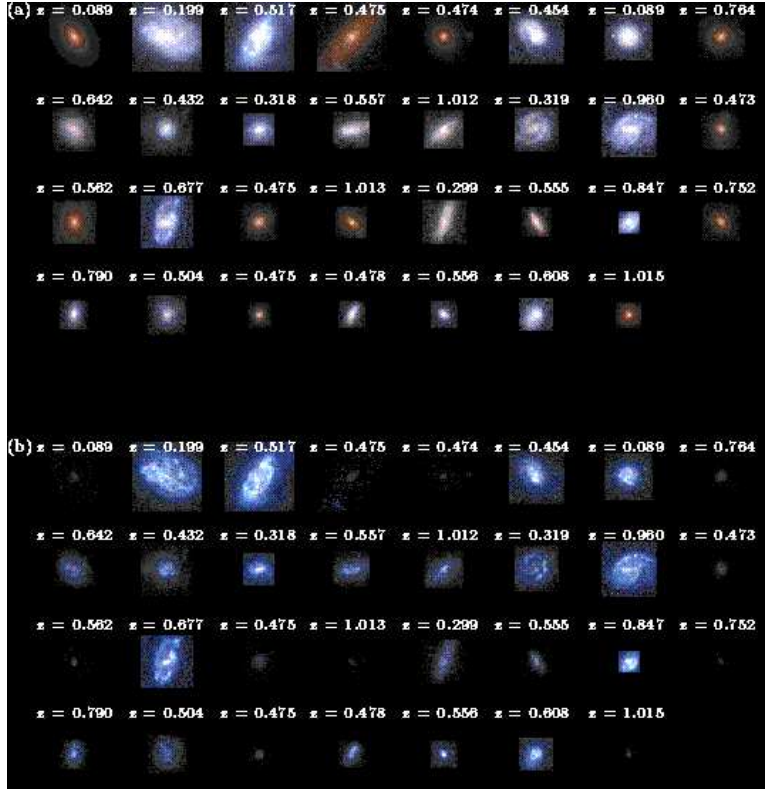}}
\end{center}
\caption{ Colour images of our magnitude-limited ($I_{814,AB} <
22.33$) redshift sample from the HDF generated from the $I_{814}$,
$V_{606}$, and $B_{450}$ images determined by k-correcting each pixel to
$z=0.5$ (panel (a)) and to $z=2.5$ (panel (b)). The intensity is
scaled to the peak surface brightness to bring out the spatial detail
at $z=0.5$ and to give a qualitative assessment of the importance of
the k-correction for various portions of the galaxies, thereby
illustrating the strong contrast between ellipticals and HII regions
at high redshift.  The ordering in this figure is by apparent
magnitude with the brightest in the upper left-hand corner and the
measured redshifts being listed for identification.  For the sake of
clarity, these images are not scaled or rebinned in area as in our
simulations.\label{stamps}}
\end{figure}

\placetable{tbl-1}
\placefigure{samplez}
\placefigure{stamps}

\section{Representation of Each Sample Galaxy}

\subsection{The Light}

We treat our bright galaxies as two-dimensional pixelated
light-emitting surfaces for the purpose of k-correction, avoiding the
unnecessary step of parameterizing their light profiles. We do not
attempt to alter the inclination of each galaxy in the simulations,
with the attendant problems regarding extinction this would entail,
since by default our sample already contains galaxies of all
orientations.

The question arises as to the best method of k-correction when the
redshift extends to wavelengths short enough that no UV information is
available.  The spectrum of any given pixel is only sampled by 4 broad
passbands and no information exists shortward of 2000 $\AA$ in the
restframe of the typical bright galaxy (for which $\bar{z}\sim0.5$),
allowing a reliable k-correction to $z\sim3$ in the longest wavelength
band, $I_{814}$.  In order to extend this work to higher redshift and
in order to implement a smooth interpolation between the four
passbands we adopt empirical template spectra.  To indicate the
robustness of the results to the choice of templates, we compare
simulations for two independent sets of templates, one based on the
small but large aperture UV sample of Coleman, Wu \& Weedman (1980)
and the other on the larger but small aperture sample of template
spectra carefully compiled and combined from the IUE archive by Kinney
et al.\ (1996).  These data sets provide useful UV spectra for a range
of optically selected galaxies nicely spanning the observed spectral
range (see Connolly et al.\ 1996), so that a reasonable empirical
interpolation and extrapolation can be performed by matching these
templates to the 4 passbands of the HDF and improved by interpolating
between these spectra to provide a smooth result.  Additionally, we
include the average effects of the HI Lyman continuum and series
forest absorption at high redshift, as parameterized by Madau (1995)
for distant QSO's, since they significantly affect the broad-band
colours of high redshift galaxies.

Formally, we represent each element $(\vec{x})$ in the two-dimensional
template of a galaxy by
\begin{equation}
F_{\vec{x}} (\lambda) = \mu(\vec{x}) f_{s(\vec{x})} (\lambda)
\end{equation}
where $\mu(\vec{x})$ is the surface brightness template at element
$\vec{x}$ in the $b$ band and where $f_{s(\vec{x})} (\lambda)$ is the
spectral energy distribution shape at the element $\vec{x}$.  We take
the surface brightness template to equal that observed in the HDF in
band $b$, the band with the highest integrated signal-to-noise (either
the $I_{814}$ or the $V_{606}$ band for the objects in our sample.)
For clarity we use the word ``element'' when referring to a pixel in
the original bright galaxy image because these pixels are transformed
in area and flux by redshift, and therefore distinct from the pixel
scale of the simulations which is set by the HDF.

We take the SED shape, $f_{s(\vec{x})} (\lambda)$, for each element
$\vec{x}$ from two different compilations of spectral templates: the
Coleman, Wu, \& Weedman (1980) (hereinafter, CWW) set supplemented
with the NGC4449 spectrum (Bruzual \& Ellis 1985, unpublished) and the
Kinney et al.\ 1996 (K96) spectra.  CWW includes SEDs intended to be
representative of the dust-free E, Sbc, Scd, Sdm, and starburst
(NGC4449) galaxies, and the K96 set includes SEDs intended to
represent ellipticals, Sa, Sb, and SB1 (starburst; $E(B-V) \leq 0.3$)
galaxies.  We extend these observed SED templates below 1200 $\AA$ and
1400 $\AA$ respectively, by extrapolating the slope of the SED at
these wavelengths down to the Lyman break at 912 $\AA$, below which we
set the observed flux to zero (see Figure 3).  We also interpolate
linearly between spectra to form a smoothly continuous set for better
fitting the observed 4 passband flux measurements.  

Integrating the SEDs over the transmission curve of the 4 passbands,
we then find the most-likely spectral template for each element
$\vec{x}$, which we shall call $s(\vec{x})$, such that the sum of the
squares of the differences between the SED template fluxes and the
observed fluxes divided by the expected error in these fluxes is a
minimum.  The expected error is taken to be the error in the value of
the pixel fluxes added in quadrature with the typical cosmic variance
in the model SED fluxes ($0.^m25$) we find.  Note that where the
signal in a pixel is less than twice the sky noise, we have set the
signal in that pixel equal to the value given by Eq. (1), adopting the
surface brightness profile and the spectral energy template which best
matches the mean value determined in a 1.0 arcsec diameter aperture.
We have not attempted to realign the HDF images in the different
bands, nor have we tried to correct for the different shapes that the
PSF has in the different bands, since checks indicate that these
effects are small and induce changes on scales smaller than the HDF
pixel scale, i.e., 0.04 arcsec.  This wavelength independence is due
to the undersampling of images by WFPC, the effective HDF PSF being
more set by the wavelength independent sub-sampling grid pattern than
by the wavelength dependent diffraction of the telescope.

\begin{figure}
\epsscale{1.00}
\plotone{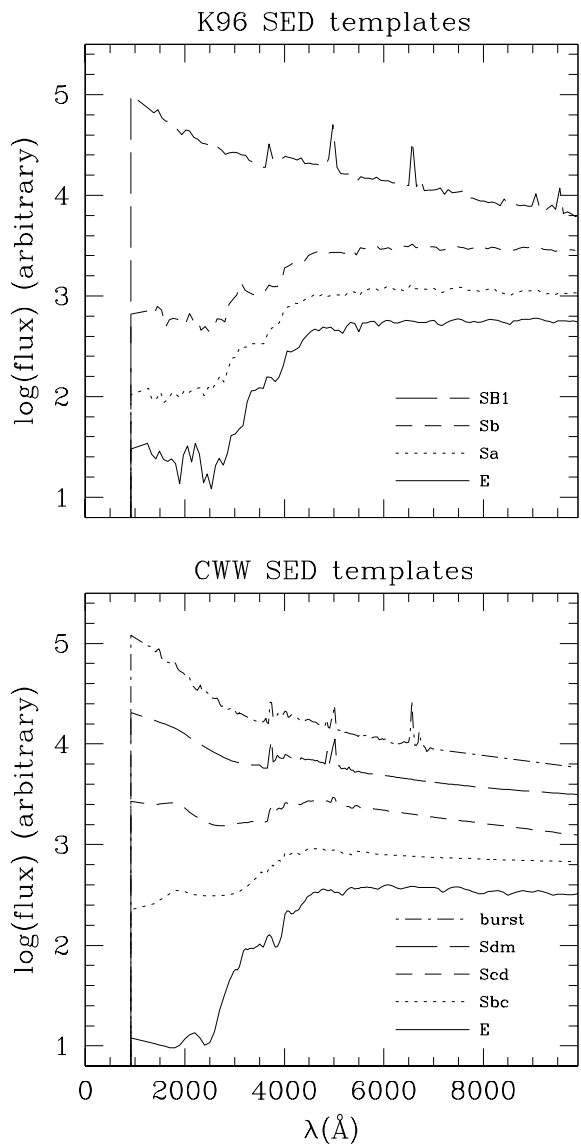}
\caption{The two spectral energy distribution (SED) template sets used
for the purposes of pixel-by-pixel k-correcting our bright HDF sample.
The upper panel shows the set compiled by Kinney et al.\ (1996) and
the lower panel shows the set compiled by Coleman, Wu, \& Weedman
(1980) which we have augmented with the NGC4449 spectrum (Bruzual \&
Ellis 1985, unpublished).  In practice, the best-fit linear
interpolation between these template spectra are adopted for the
purposes of k-correction.\label{sedtemp}}
\end{figure}

\placefigure{sedtemp}

Clearly, we should take the spatial extent of each prototype image to
be as large as possible within the limits of the signal.  In practice,
we take this to be the radius at which the mean pixel signal within an
elliptical annulus is equal to the sky noise.  Inspection of each
image allows us to filter out the small number of obviously unrelated
galaxies within this extended aperture, the pixels affected being
replaced by their reflected counterparts.  To be sure of including the
``whole'' object we extend its 2D profile by 50\%, using the radial
gradient. For this extended region we produce the noise by adding
Poisson noise from modeled signal in quadrature with the general
background noise. Except for the very brightest ``cloned'' galaxies,
this extrapolation is not relevant, since the vast majority of
replicated images generated from a given template galaxy are much
fainter than the original, lying at higher redshift where these outer
regions of the profile lie further into the noise.

The question arises as to the utility of the HDF $U_{300}$ images for
generating the higher redshift images in the redder bands, since the
sensitivity of the redder bands of the HDF is greater than that of the
$U_{300}$ band.  As it turns out, the signal in the HDF $U_{300}$
images is more than sufficient for our purposes, due to the strong
$(1+z)^4$ cosmological surface brightness dimming.  To see this,
consider the S/N at a given $\sigma$ in a pixel through the $U_{300}$
filter at the observed redshift.  This translates into a lower S/N at
higher redshift in the other bands, given their relative sensitivities
and exposure times.  Taking the worst case that the spectrum of a
given pixel is flat in frequency (though in practice the spectral
indices of some starburst systems are somewhat steeper than this), the
surface brightness dims by a factor $7.5 \log
\frac{(1+z)}{(1+z_{obs})}$, or $7.5 \log
\frac{\bar{\lambda_b}}{\bar{\lambda_{F300W}}}$, or a decrease of the
surface brightness by 1.3, 2.3, and 3.3 $\textrm{magnitudes/pixel}^2$,
for the $B_{450}$, $V_{606}$ and $I_{814}$ bands respectively.
However, the measured 1-$\sigma$ $U_{300}$ noise level for the HDF is
32.4 $\textrm{mag/pixel}^2$, compared to the 32.2, 34.0, and 32.6
$\textrm{mag/pixel}^2$ levels in the $B_{450}$, $V_{606}$, $I_{814}$
bands.  Clearly, then, since few pixels are as blue or bluer than a
flat spectrum in our template galaxy sample, the S/N of any pixel in a
redder band at higher redshift will always be less than that of the
observed $U_{300}$ image.

While it is apparent that a number of objects in our input sample have
extremely low values of S/N in the $U_{300}$ band (see the ellipticals
in Figure 2a), we emphasize that the S/N in $U_{300}$ is {\it always}
sufficient to determine the appearance of these galaxies at higher
redshifts in the redder bands of the HDF given the relative band
sensitivities and exposure times used in these passbands. Greater
inequities between bands would limit the simulations to depths less
than the limiting magnitudes of these passbands.

\subsection{Space Density}
					
We set the space density of each galaxy in our sample equal to
${1/V_{max}}$, determined by the maximum redshift, $z_{max}$, to which
each galaxy could have potentially been selected given our chosen
magnitude limit.  $V_{max}$ is given by
\begin{equation}
V_{max} = S \int _0 ^{z_{max}} \left(\frac{d_L}{1+z} \right)^2 
\frac{cdz}{H_0 E(z)}
\end{equation}
where 
\begin{equation}
E(z) = \sqrt{\Omega(1+z)^3+(1-\Omega-\Omega_{\Lambda})(1+z)^2+
\Omega_{\Lambda}}
\end{equation}
where $S$ is the solid angle of the surveyed area and where $d_L
(z_{max})$ is the well-known luminosity distance at redshift
$z_{max}$.  $z_{max}$ is determined by:
\begin{equation}
22.33 = I_{814,AB} + k_{814} (z_{max}) - k_{814} (z_{obs}) + 5 \log
\frac{d_L(z_{max})}{d_L(z_{obs})}
\end{equation}
where $z_{obs}$ is the observed redshift and $k_{814}(z)$ is the
$k$-correction in the $I_{814}$ band at redshift $z$.  We determine
the $k$-correction for the galaxy from the total spectral energy
distribution, which incorporates the contribution of all the elements
in our two-dimensional representation of each galaxy.  Note this
procedure avoids the usual steps of constructing a luminosity
function, parameterizing it, and then assigning k-corrections.  We
simply treat each galaxy as a class of its own with a space density
set by its own value of $1/V_{max}$, allowing a fully unbinned,
individual treatment of each template galaxy.

\section{Consistency Checks and Sample Fairness}

  Astronomers have long used the $V/V_{max}$ distribution (Schmidt 1968)
to assess the completeness and uniformity of a sample, where $V$ is
the volume up to and including the redshift of the galaxy in question
and $V_{max}$ is the volume in which this galaxy could have been
observed, given the selection criteria.  For a uniform distribution of
galaxies in a volume one expects a uniform distribution of $V/V_{max}$ 
between 0 and 1, in the absence of evolution and clustering, and 
for a sample with $N$ galaxies the average value of this 
quantity us given by:
\begin{equation}
\left\langle \frac{V}{V_{max}} \right\rangle = 0.5 \pm
\frac{1}{\sqrt{12N}}
\end{equation}
The distribution of $V/V_{max}$ for our template sample is shown in
Figure 4, for $\Omega = 0.1$/$\Lambda = 0.9$, $\Omega = 0.1$, and
$\Omega = 1.0$, where the k-correction is formed from the SED averaged
over the whole object.  The average value for the $V/V_{max}$
distribution is 0.48, 0.51, and 0.53 for the $\Omega = 0.1$/$\Lambda =
0.9$, $\Omega = 0.1$, and $\Omega = 1$ geometries, which is within the
expected deviations for the $V/V_{max}$ statistic $(0.50 \pm 0.05)$
and not inconsistent with the rate of evolution detected in a larger
redshift survey to a similar magnitude limit (Lilly et al.\ 1995).  We
could have chosen to derive $V_{max}$ by redshifting each pixel of the
object and performing the photometry on the redshifted image for
greater self consistency, but this turns out to be virtually
irrelevant because the magnitudes we recover from the redshifted
images placed at $z_{max}$ are on average displaced by only 0.07
magnitudes faintward of our magnitude limit ($I_{814,AB} = 22.33$).  A
slight faintward shift is expected for aperture magnitudes since the
lower surface brightnesses at $z_{max}$ results in some small loss of
the light on the wings.  To illustrate this point, we compare the
recovered magnitudes at $z_{max}$ with the magnitude limit chosen for
our bright sample in Figure 5.

\begin{figure}
\epsscale{1.00}
\plotone{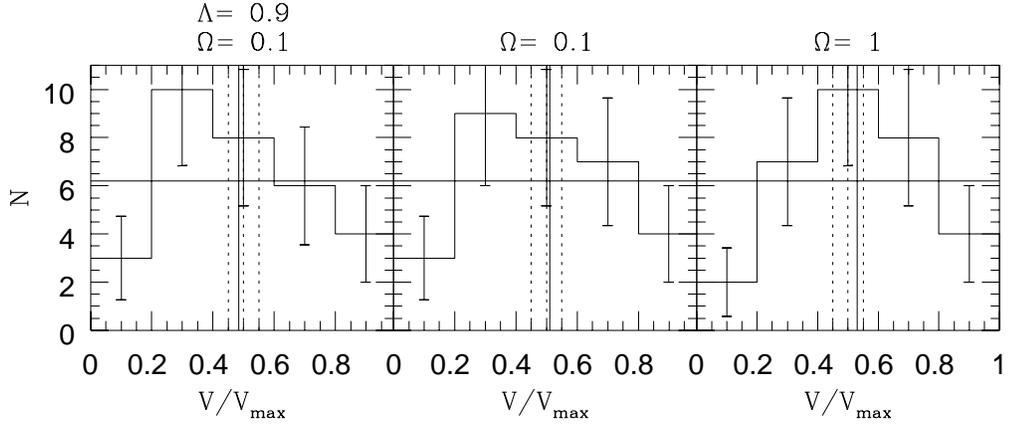}
\caption{The $V/V_{max}$ distribution for our sample of galaxies with
$1\sigma$ Poissonian error bars determined for the $\Omega =
0.1$/$\Lambda = 0.9$, $\Omega = 0.1$, and $\Omega = 1$ geometries.
The solid vertical line indicates the average value of $V/V_{max}$,
and the dotted vertical lines bracket the one sigma range for a
homogeneous sample of 31 galaxies.  The bottom panel shows this for
the ``no-evolution'' case.  Note that for our faint galaxy simulations
the calculation of the sample galaxy densities, given by $1/V_{max}$,
are made self-consistently.  This figure shows that the bright sample
is evenly spread in the interval $0<V/V_{max}<1$ given the errors, for
any choice of geometry, indicating we have a fair representation of
galaxies.\label{vvmax}}
\end{figure}

\begin{figure}
\epsscale{0.95}
\plotone{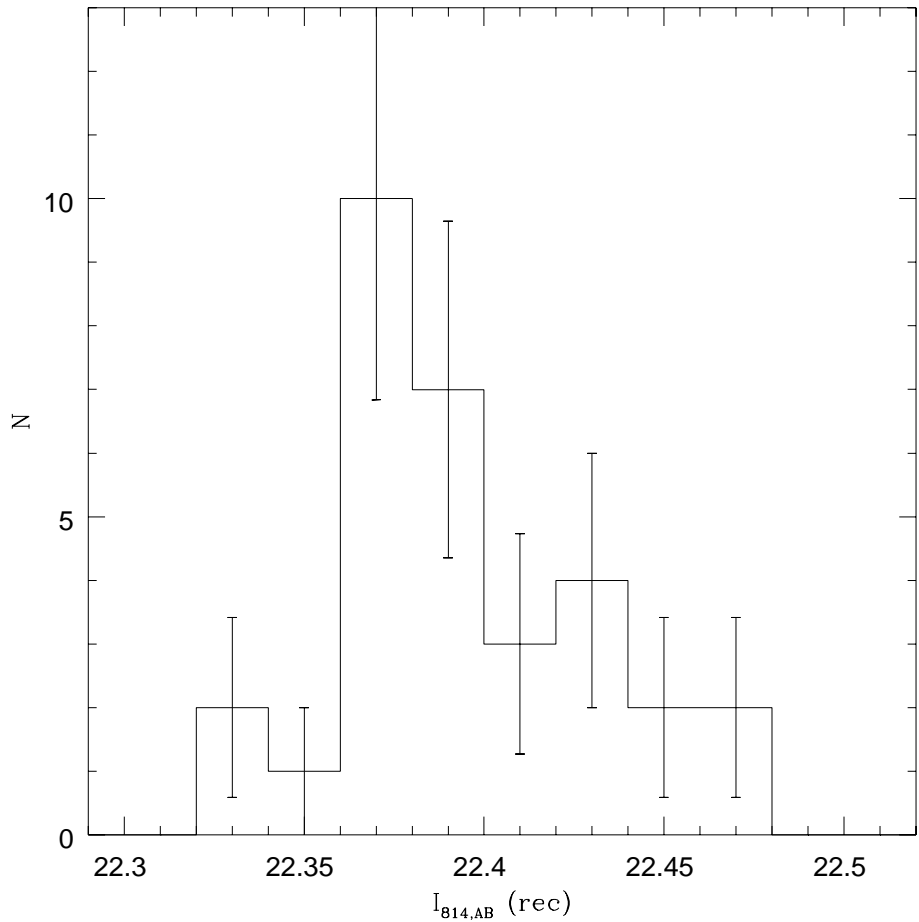}
\figcaption{The recovered magnitudes of our sample galaxies after placing
them at their values of $z_{max}$, corresponding to the magnitude
limit of the data $I_{814,AB} = 22.33$.  $z_{max}$ is determined using
the integrated light, whose mean SED was taken as an integral over all
the pixels.  This plot shows that the determination of $z_{max}$ (and
hence the space density $1/V_{max}$) derived in this way is very close
to the value that would be determined using two-dimensional
k-corrections, albeit slightly biased, since the recovered magnitudes
are only 6\% fainter than the magnitude limit.  That the recovered
magnitudes are systematically on the faint side of this limit is to be
expected since a portion of the outer light profiles are lost in the
sky noise.\label{vmaxm}}
\end{figure}

\placefigure{vvmax}
\placefigure{vmaxm}

To provide a basic context for understanding the no-evolution
simulations described above we display the luminosity function
obtained for this sample in Figure 6, for two different representative
cosmologies.  We compare this luminosity function with the $b_j$-band
luminosity functions determined by Loveday et al.\ (1992) and Zucca et
al.\ (1997) in the APM and ESP surveys, respectively correcting our
sample to restframe $b_j$ directly using our pixel-by-pixel best-fit
SEDs.  We see that our luminosity function is shifted to larger
luminosity and/or space density than that of the local universe,
broadly consistent with the findings of Ellis et al.\ (1996) and Lilly
et al.\ (1995) where the luminosity function of blue objects is
observed both to brighten and to steepen.

\begin{figure}
\epsscale{0.95}
\plotone{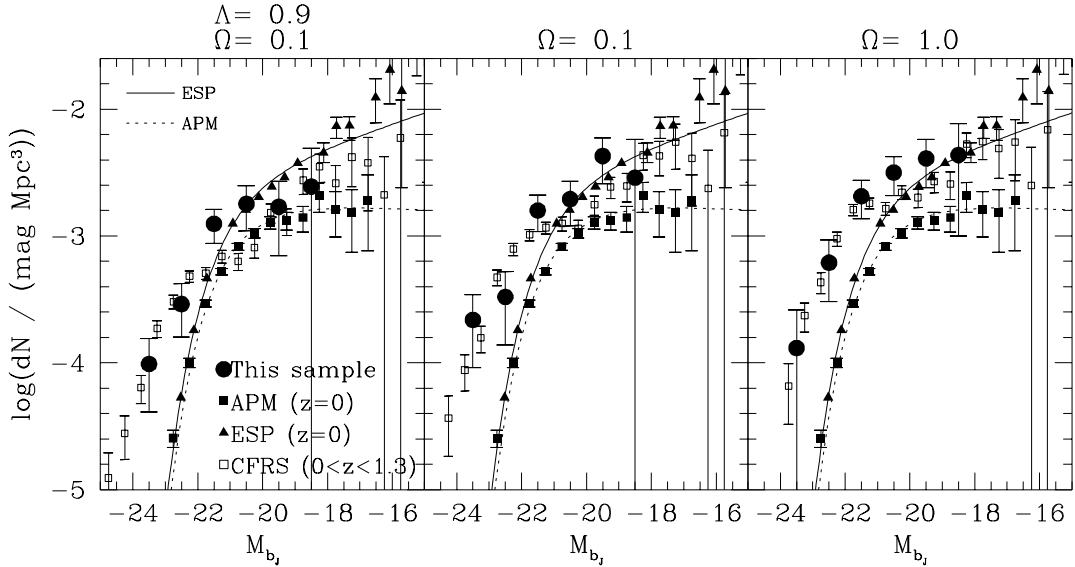}
\figcaption{Luminosity functions determined for the galaxies in our
sample (solid circles) colour transformed to the $b_j$-band, by using
the integrated SED determined from the pixel-by-pixel fits to the
broadband colours.  We compare these luminosity functions to the local
luminosity functions of the APM survey (Loveday et al.\ 1992) and the
ESP $b_{j}$-band luminosity function of (Zucca et al.\ 1997) plotted
as solid squares and triangles, respectively.  All error bars
represent $1 \sigma$ Poissonian uncertainties.  Also shown is the
transformed CFRS luminosity function to the $b_{j}$-band using their
published $V-I$ colours and the CWW SED templates for interpolation.
Note the fairly good agreement of our luminosity function with the
CFRS and that both of our luminosity functions show evolution with
respect to the APM and ESP.\label{lf}}
\end{figure}

\placefigure{lf}

We can also compare our sample with the redshift distribution, N(z),
of the Canada-France Redshift Survey (CFRS), a survey of similar
depth, by simulating (as described in detail below) its survey
parameters.  The CFRS sample covers $17.5 < I_{AB} < 22.5$ over 112
$\textrm{arcmin}^2$.  Scaling by these criteria and the 19\% redshift
incompleteness of the CFRS, we can construct a prediction for N(z)
using our method (Figure 7).  We make a small correction, inferable
from Figure 5 of Lilly et al.\ (1995), to convert from isophotal to
total magnitudes by a uniform 0.1 mag.  A noticeable difference
between our predicted N(z) and that of the CFRS is found in the
amplitude of the distributions, our redshift distribution being 28\%
higher, consistent with the differences in the respective luminosity
functions (Figure 6).  Despite this, the shape of our predicted N(z)
is very similar to that of the CFRS, which is not surprising given the
wide spread of $V/V_{max}$ for our sample (Figure 4).  This basic
agreement is quite satisfactory, showing that our bright galaxy sample
fairly samples the range, if not quite the average density, of
galaxies comprising the general field population.

\begin{figure}
\epsscale{0.95}
\plotone{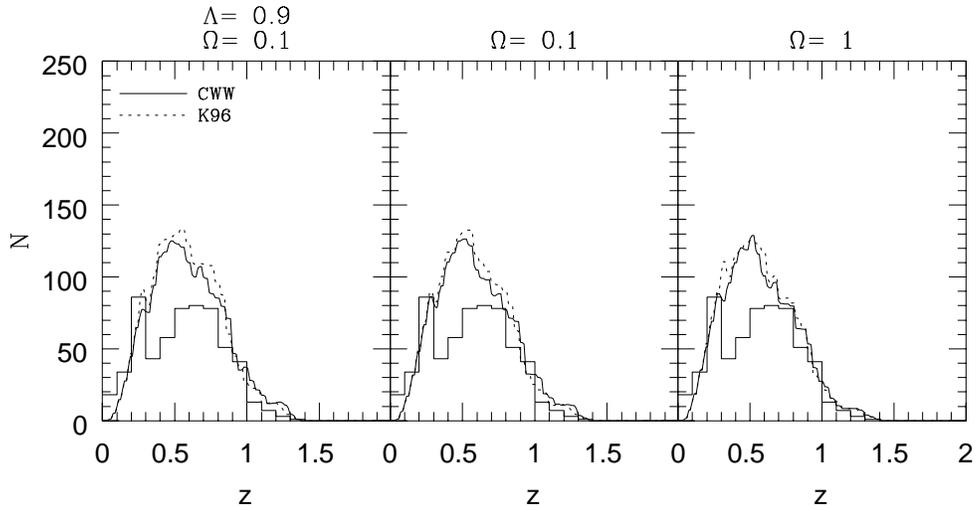}
\caption{Comparison of the redshift distribution for the CFRS
(histogram) with that predicted from our simulations for $17.5 <
I_{AB} < 22.5$ for $\Omega = 1$, $\Omega = 0.1$, and $\Omega =
0.1$/$\Lambda = 0.9$ using two different sets of SED templates (solid
lines = CWW SED templates; dotted lines = K96 SED templates) for the
interpolation between the HDF bands.  The comparison accounts for the
relative areas of the two redshift samples and assumes the 19\%
incompleteness in redshift of the CFRS as just a normalization
correction.  The HDF bright sample lies 28\% higher in number than the
CFRS, almost certainly reflecting the clustering seen in the redshift
distribution (see figure 1). Note that the shapes of the distributions
are very similar, suggesting that our sample has a representative mix
of galaxies.\label{dndz}}
\end{figure}

\placefigure{dndz}

\section{HDF Simulations}

According to the number density derived for each template galaxy,
$1/V_{max}$, and assuming homogeneity, we generate Monte-Carlo
catalogues of the objects out to very large redshift ($z=7$) over the
solid angle of the HDF, each object being assigned a random redshift,
position, and position angle.  We then generate mock images on the
basis of these catalogues.  Each redshifted image must be scaled in
size and resampled with more noise and additional PSF smoothing to
account for its higher redshift and generally smaller size relative to
its brighter counterpart.  Since the two-dimensional surface
brightness profiles of the bright galaxies already contain noise, we
find it useful to simultaneously generate both a signal $I$ and noise
$N$ image.  The noise image $N$ keeps track of how much noise has been
implicitly added to each pixel in the signal image $S$ by virtue of
each galaxy template implicitly containing noise.  Such an accounting
allows the proper amount of noise to be added to each pixel in the
signal image after all the scaled galaxy templates have been laid
down.

To generate an image at a chosen redshift, we must
calculate the change in size and surface brightness of each element
$\vec{x}$ in our two-dimensional galaxy templates.  For a galaxy with
redshift, $z$, we take the angular size of each element in our
two-dimensional galaxy template, $d_z$, to be equal to
\begin{equation}
d_{z} = \left ( \frac{d_A(z_{obs})}{d_A(z)} \right ) d_{gal}
\end{equation}
where $z_{obs}$ is the redshift of the object as measured in the HDF,
$d_A$ is the angular distance (Peebles 1993), and $d_{gal}$ is the
angular size of each element in our two-dimensional galaxy template
image at $z_{obs}$, i.e. the pixel size of galaxies in the HDF (0.04
arcsec).  We calculate the surface brightness $\mu^X(\vec{x},z)$ that
each element $\vec{x}$ of our two-dimensional galaxy template has at
redshift $z$ for a given passband $X$ as
\begin{equation}
\mu^X(\vec{x},z) = \mu(\vec{x}) + k^X(\vec{x},z) - k^b (\vec{x},z_{obs})
 + 2.5 \log \left ( \frac{1+z}{1+z_{obs}} \right) ^4 + (X
 - b)(\vec{x})
\end{equation}
where $k^b(\vec{x},z)$ and $(X - b)(\vec{x})$ are the $k$-corrections
and $z=0$ $X - b$ colours calculated based on the spectral energy
distribution $f_{s(\vec{x})}$ of element $\vec{x}$.  We have included
the line blanketing by the Lyman-alpha forest as given by Madau (1995)
as well as Lyman-limit absorption in our calculations of the
$k$-corrections since these corrections become important at $z>2.0$ in
the $U_{300}$ band and at $z>3.5$ in the $B_{450}$ band.

In calculating both the signal and the noise at each pixel on the
images $I$ and $N$, we break up each pixel into 25 smaller subpixels,
and calculate the contribution of each redshifted element $\vec{x}$ to
each of these smaller subpixels to account for the generally much
larger area covered by a data pixel than that of the redshifted image
element, for the purpose of rebinning.

Because the real PSF has smoothed each galaxy template, to correctly
calculate the appearance of each prototype galaxy in our sample at
higher redshifts we must add more smoothing present in each object in
the $I$ image, depending on the reduction in angular size for the
object in question.  For simplicity, we simply smooth each object with
a kernel derived from a relatively isolated, unsaturated star from the
Hubble Deep Field reduced in size so that its scale length is simply
$\sqrt{1 - \left (\frac{d_A(z_{obs})}{d_A(z)}\right )^2}$ times that
of the original scale length.  The above expression is exact for the
case of a perfectly Gaussian PSF and is also close to exact in those
cases where the angular size of the simulated galaxy laid down in the
$I$ image is much smaller than the HDF.  Of course, it is true that
the real HDF PSF differs from a Gaussian in that it has much more
extended wings, but for the most part the differences that this makes
are small.  To verify this, we compared the angular sizes recovered
using the present procedure and from assuming the PSF to be exactly
Gaussian (the $\sigma$'s for which we determined by fitting to the
same unsaturated star in the HDF), and we found little if any
dependence on these differences.

For a pixel in which the contribution to the signal image $I$ from an
element in the template image is $I(\vec{x})f$, the noise contribution
to the same pixel in the noise image $N$ is taken to be
\begin{equation}
\frac{I(\vec{x})f}{(\frac{S}{N})(\vec{x})} \sqrt{\frac{d_{z}}{d_{gal}}}
\end{equation}
added in quadrature, where $(\frac{S}{N})(\vec{x})$ is the S/N ratio
calculated for the element $\vec{x}$ for the $\mu(\vec{x})$ template
drawn from the HDF, and account is made of both the background and
Poissonian noise associated with the measured pixel-by-pixel signal.

Having generated the signal image, we want to calculate the
appropriate noise for each pixel in our simulated HDF, which we take
to be:
\begin{equation}
N(\vec{x})_{desired} = \sqrt{\sigma^2 + (I(\vec{x}))^2 G}
\end{equation}
where $\sigma$ is the background noise level in the HDF, $I(\vec{x})$
is the signal at pixel $\vec{x}$, and $G$ is the gain.

Having already calculated the amount of noise which each pixel in the
signal image had by virtue of implicit noise in the templates, i.e.
the noise image $N$, we can bring the noise in each pixel up to the
appropriate value $N_{desired}$ by adding pixel-by-pixel
Gaussian-distributed noise with standard deviation
$\sqrt{N_{desired}^2 - N^2}$, smoothing this added noise with the
noise kernel specified in Table 4 of Williams et al.\ (1996) so as to
approximate the observed correlation properties of the noise in the
HDF.  While the outlined procedure will add the appropriate amount of
noise to pixels if noise is lacking in those pixels, it is possible
that some pixels will already have more noise added to them than is
present per pixel in the HDF.  In particular, this occurs when the
redshift for an object is lower than that of the prototype, since of
course we cannot improve on the S/N of the prototype galaxy image.  To
artificially prevent this possibility, we simply replace every galaxy
in our mock catalogue whose redshift is lower than its observed
redshift, with another galaxy from our bright sample placed at the
same redshift whose observed redshift is lower than the mock catalogue
redshift.  Though this results in a bright galaxy population which is
slightly unrepresentative (and, in fact, this is the cause of the
slight disagreement observed in the distribution of bright HDF
galaxies and the simulations found later in the paper), it does not
appreciably bias the measured properties of the cloned galaxies at the
fainter magnitudes of interest.

The simulations cover a sky area four times that of the HDF in each of
the four broadbands $U_{300}$, $B_{450}$, $V_{606}$, and $I_{814}$.
We perform these simulations self-consistently for $\Omega =
0.1$/$\Lambda = 0.9$, $\Omega = 0.1$, and $\Omega = 1$, by which we
mean that the $V_{max}$ of the bright sample uses the same geometry as
the volume used in constructing the simulation.  In Figure 8 we
compare a no-evolution simulation assuming $\Omega = 1.0$, $\Omega =
0.1$, and $\Lambda=0.9$/$\Omega=0.1$ with an area of the same size
selected from the HDF.

\begin{figure}
\begin{center}
\resizebox{11.5cm}{!}{\includegraphics*[125,130][500,650]{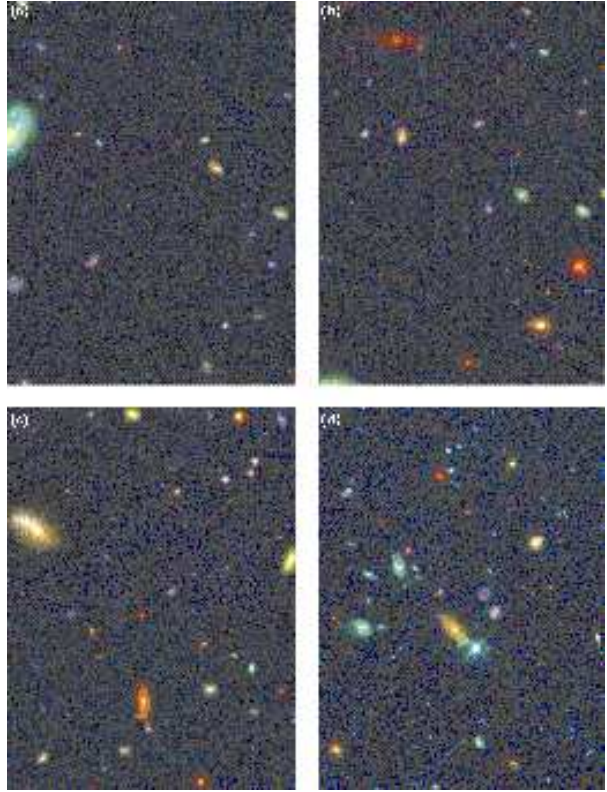}}
\end{center}
\caption{Panels (a), (b), and (c) show 52'' x 72'' colour images
generated from our no-evolution simulations for the $\Omega = 1.0$,
$\Omega = 0.1$, and $\Lambda = 0.9$/$\Omega = 0.1$ geometries,
respectively, of the $B_{450}$, $V_{606}$, and $I_{814}$ images,
constructed with pixel size, S/N, and PSF identical to that of the
HDF.  Panel (d) shows an image of the same size taken from the HDF.
Clearly, the no-evolution simulation strongly underpredicts the total
number of faint galaxies in the HDF.\label{rne}}
\end{figure}

\placefigure{rne}

A simple test of our procedure is to construct images of the 31
prototype galaxies at their observed redshifts ($z_{obs}$).  Good
agreement is found between these images and their originals at
$z_{obs}$, within the errors expected on the basis of our
pixel-by-pixel SED fits.  To illustrate this agreement more
quantitatively, we plot a comparison of the recovered values for the
half-light radius, Petrosian radius (Petrosian 1976), and apparent
$I_{814}$ magnitudes in Figure 9.

\begin{figure}
\epsscale{1.0}
\plotone{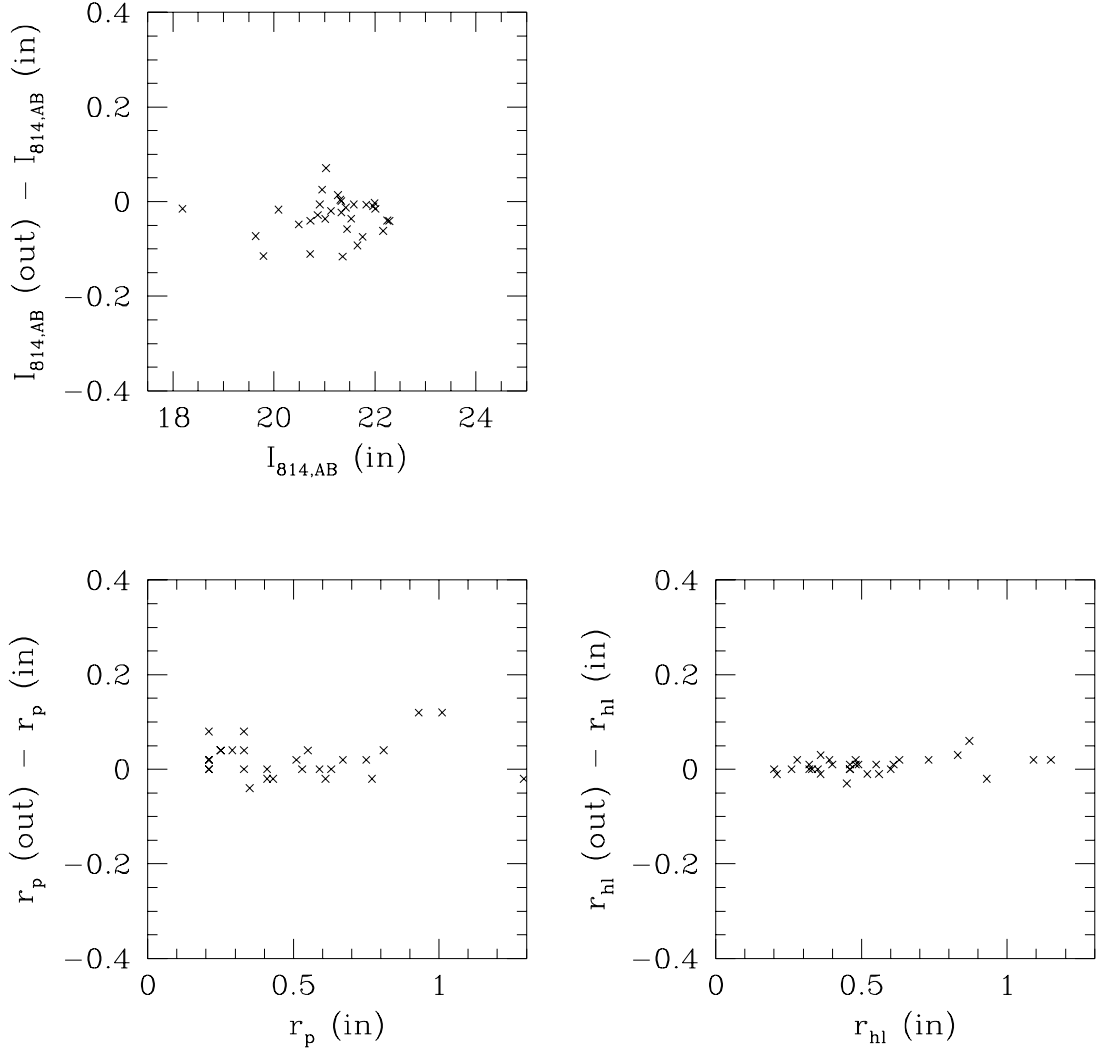}
\caption{A comparison of the apparent magnitudes, Petrosian radii, and
half-light radii extracted from the HDF $I_{814}$ images against those
sample bright galaxies laid down on a simulated image at their
observed redshifts for our bright sample of galaxy prototypes.  Note
that the recovered values are very close to the true observations,
thereby validating our technique of image generation.\label{rrec}}
\end{figure}

\placefigure{rrec}

\section{Object Detection}

On both the simulations and the HDF itself, we perform the object
identification and photometry using SExtractor version 1.2b5 (Bertin
\& Arnouts 1996).  After smoothing the $I_{814}$ images with a
Gaussian of 0.06-arcsec radius (the approximate PSF) within
SExtractor, we require that objects be at least 2 sigma above the
noise over an area of at least 10 contiguous pixels, and we use a
cleaning parameter of 1.0.  We select our apparent magnitudes to be
equal to SExtractor's $MAG\_BEST$ estimate, which for non-crowded
objects is equivalent to an isophotal magnitude extrapolated beyond
the isophotes.  We exclude objects with half-light radii less than
0.15 arcsec and $I_{814,AB} < 23$ as stars, fainter than which the
contamination is relatively small.  Then, based upon the detected
objects, we used SExtractor to determine the apparent magnitudes in
the other bands with the $I_{814}$ apertures.

 We set the deblending parameter ($DEBLEND\_MINCONT$), which is
important in determining the extent to which SExtractor breaks up
objects, to be equal to 0.04.  We found some dependence of our results
on the value chosen for this parameter, particularly the break-up
rate, but for the most part this dependence was small.  Our chosen
value of the deblending parameter is very close to that (0.05) used by
Clements \& Couch (1997) (Couch, private communication).  With this
choice of parameter, we obtained the reasonable result that no bright
galaxy from our sample broke up into smaller pieces when placed at its
$z_{max}$, in agreement with our qualitative
impressions in looking at these images.

We derive two different measures of the angular size for each object
in our image, a Petrosian radius (Petrosian 1976) and a half-light
radius.  We take the Petrosian radius to equal the smallest radius for
which the surface brightness at that radius equals half the average
surface brightness interior to that radius.  The half-light radius is
equal to the radius of the aperture which contained half the light as
determined by SExtractor's best estimate of the total light.  We
performed some simulations to test our method for recovering
half-light radii and found that our recovered half-light radii are
only slightly scattered about sizes 10\% smaller than the input
half-light radii at $I_{814,AB} \sim 24$.  The scatter in this
relationship partially derives from the uncertainty in the overall
photometry performed by SExtractor.

\section{Results}

\subsection{Number Counts in the $I_{814}$ band}

It is natural to begin our comparison of the faint galaxy population
in the HDF by looking at the counts in $I_{814}$, the longest
wavelength band and hence least affected by uncertainty in the
k-correction. Figure 10 compares the number counts recovered from both
our simulations and the observations (a description of the plotted $1
\sigma$ estimates based on the size of our bright sample is given in
Appendix A.)  For all geometries, we find that our no-evolution
predictions fall steadily short of the observations with increasing
apparent magnitude.  For $\Lambda=0.9$/$\Omega=0.1$, the shortfall is
noticeably less pronounced due to the relatively larger volume
available.  At $I_{814,AB} = 26$, for example, the counts fall short
by a factor $1.9\pm 0.4$ for $\Lambda=0.9$/$\Omega=0.1$ compared to
$5.4\pm 1.0$ for $\Omega=1$ and $2.7\pm 0.8$ for $\Omega=0.1$.

Of course, the counts recovered at bright magnitudes, $I_{814,AB}
\approx 22$, are in fair agreement with the observations, as one would
expect given the definition of our sample up to approximately the same
bright limit, $I_{814,AB} = 22.33$.  It is reassuring to find that
there little dependence of the number counts on the choice of spectral
template, whether it be the CWW set or the K96 set.  Finally, as
discussed in Appendix B, we note that the present no-evolution faint
counts ($I_{814,AB} \sim 28$) are compromised of a non-negligible
number of the input prototype galaxies ($\gtrsim 10$).

\begin{figure}
\epsscale{0.9}
\plotone{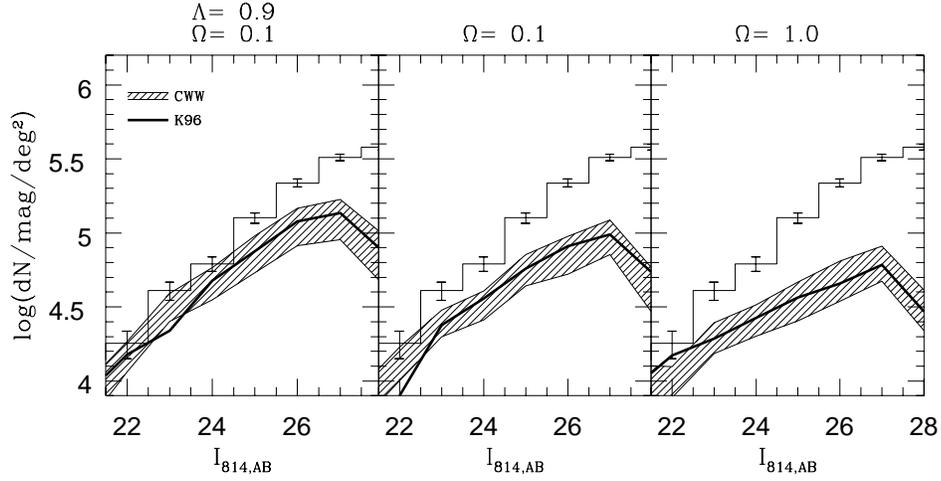}
\caption{The upper panels show the comparison of the observed number
counts (histogram with $1 \sigma$ Poisson errors) in $I_{814,AB}$ with
those recovered from our ``no-evolution'' simulations for the two
choices of SED templates.  The hatched region represents the estimated
1 $\sigma$ range in the counts based on the finite size of our bright
input sample using the CWW templates while the solid curve represents
the recovered counts using the K96 templates.  All cases are shown for
$\Omega = 0.1$/$\Lambda = 0.9$, $\Omega = 0.1$ and $\Omega = 1$
geometries.
\label{dndm}}
\end{figure}

\placefigure{dndm}

\subsection {Number-Count Completeness, Overcounting, and Clustering}

Given our knowledge of the positions, redshifts, magnitudes, and types
of galaxies laid down in each simulated image, it is simple to
determine systematic uncertainties such as incompleteness in the
number counts, by matching up the object catalogues from the
simulations created by SExtractor with our input Monte-Carlo
catalogues.  Figure 11 shows that the incompleteness becomes
significant in the range $I_{814,AB} > 26$ for both our no-evolution
simulations.  This incompleteness stems from the fact that at fainter
magnitudes, we are sensitive to smaller and hence the higher surface
brightness galaxies, so that large fraction of the population of
galaxies at faint magnitudes subtend areas too large for detection
given their redshifted surface brightnesses.

\begin{figure}
\epsscale{1.05}
\plotone{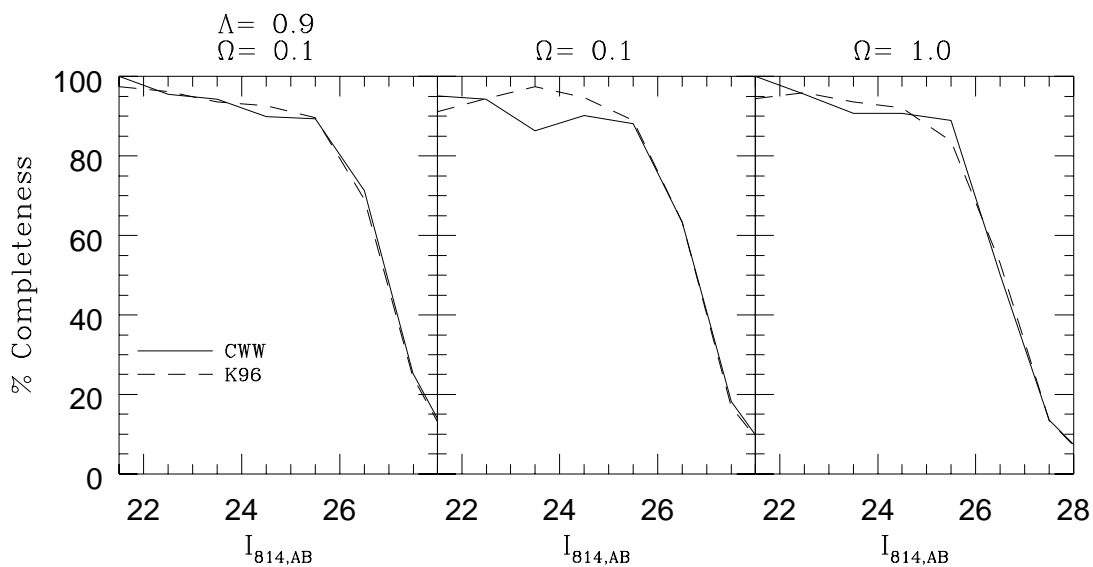}
\caption{The completeness of the counts in the $I_{814}$ band
determined from the simulations for the no-evolution simulations based
on the CWW SED templates (solid line) and the no-evolution simulations
based on the K96 SED templates for $\Omega = 0.1$/$\Lambda = 0.9$,
$\Omega=0.1$ and $\Omega=1$.  Because surface brightness has a rough
inverse proportionality to angular size at a given magnitude, the
completeness limit is directly related to the angular sizes of the
faint galaxy population in that galaxy populations with smaller
angular sizes are more complete at fainter
magnitudes.\label{complete}}
\end{figure}

\placefigure{complete}

In a similar manner, we can determine the rate of overcounting of in
our simulated fields.  As explained in Colley et al.\ (1997), it is
possible to overcount the number of faint galaxies by misidentifying
individual parts of a galaxy as distinct galaxies, especially at
higher redshifts, where the rest-frame ultraviolet light is accessed
and HII regions have a higher contrast (cf. the galaxy at $z=0.319$ in
Figure 2b).  For simplicity, rather than perform an angular
correlation analysis like Colley et al.\ (1997), we have chosen to
measure this number directly.  In Figure 12, we display the rate of
overcounting for all the simulations performed by comparison of the
input random catalogue with those recovered.  Clearly, in our
no-evolution simulations, using the parameters chosen for the
photometry, overcounting is never an important effect for any of the
cosmologies examined.

\begin{figure}
\epsscale{1.05}
\plotone{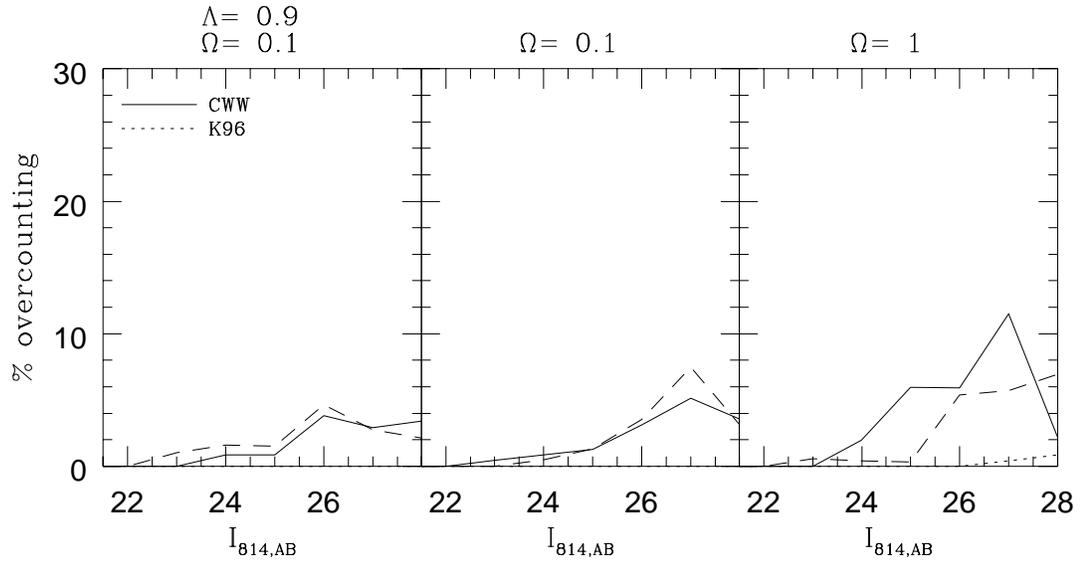}
\caption{ The \% of galaxies which are counted more than once, mostly
as a result of the fact that in the UV they break up into distinct
pieces.  The overcounting rate is relatively low and similar for all
geometries.\label{overcount}}
\end{figure}

\placefigure{overcount}

The clustering seen in the HDF redshift data at bright magnitudes
(Figure 1) is responsible for the overdensity at bright magnitudes
$I_{814}<22.33$ in the HDF relative to the mean field counts measured
in wider field surveys.  Compared to the CFRS, for example, this
amounts to a $\sim28\%$ overdensity (Figure 7).  At fainter
magnitudes, however, the count variance should be less affected by
clustering, given the greater projected volume and the expectation of
less well developed structure at earlier times.  We provide a quick
test of the plausibility of this hypothesis using other deep HST
$I_{814}$ data.  Figure 13 shows the count variance between 6
pointings (two sets of 3 HST pointings separated by 3 hours on the sky
from another program). These data show that the count variance in
$I_{814}$ band decreases steadily to the Poissonian limit ($I > 25$)
over the area of WFPC2, consistent with isotropy at faint magnitudes.

\begin{figure}
\epsscale{0.95}
\plotone{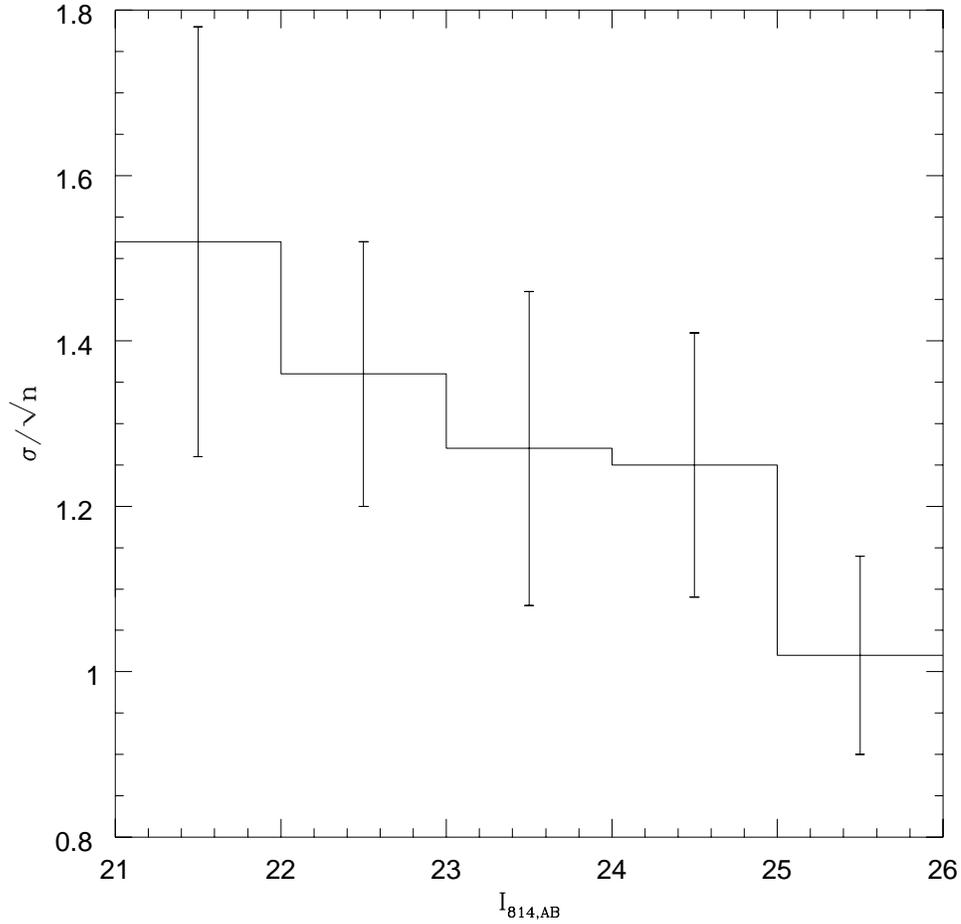}
\caption{The variance in the number counts in the $I_{814}$ band from 6
independent faint WFPC pointings from another program.  The variance
tends steadily towards the Poissonian limit.  This simple empirical
demonstration suggests that our use of the small HDF region as a fair
representation of the universe is reasonable.
\label{variance}}
\end{figure}

\placefigure{variance}

\subsection{Angular Sizes}

We compare the distributions of half-light radii recovered from the
HDF with those of our ``no-evolution'' simulations in Figure 14.  The
hatched area represents the $1 \sigma$ uncertainty range based on the
finite size of our bright sample for the no-evolution model using the
CWW SED templates.  The solid curve indicates the angular size
distribution recovered from simulations using the K96 SED templates.
At bright magnitudes ($21 < I_{814,AB} < 22.6$), the angular sizes
recovered from the simulations agree quite well with the observations
as expected given the fact that we defined our bright sample in terms
of many of these same galaxies.

\begin{figure}
\epsscale{0.91}
\plotone{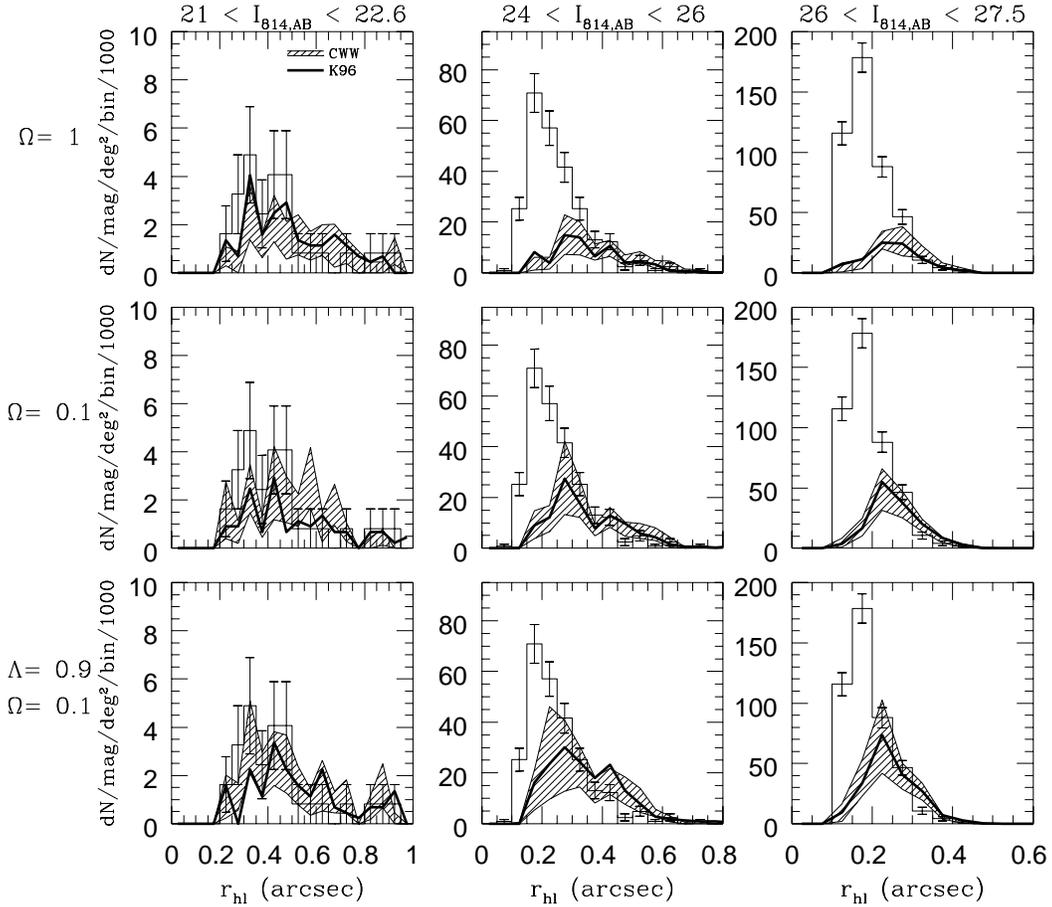}
\caption{Comparison of the distribution of half-light radii recovered
from the HDF (histogram) with those recovered from our no-evolution
simulations for three different geometries ($\Omega = 0.1$, $\Lambda =
0.9$; $\Omega=0.1$; $\Omega=1$) and three different magnitude ranges.
The hatched region represents the angular sizes recovered from the
``no-evolution'' simulations generated using the CWW SED templates
while the solid curve represents those sizes recovered from
simulations generated using the K96 SED templates.  Note the good
agreement between the recovered sizes for the no-evolution models and
the data at bright magnitudes, as required.  In contrast, for the
fainter magnitude bins, $I_{814,AB}>24$, the observed sizes are much
smaller than for the simulations.  Low $\Omega$ and the addition of
low-luminosity galaxies (see Bouwens, Broadhurst, \& Silk 1998) help
somewhat with this discrepancy, but the shortfall in number is still
large.\label{angdista}}
\end{figure}

\placefigure{angdista}

In contrast, at fainter magnitudes, the half-light radii from the
no-evolution simulations become significantly larger than those from
the observations: larger by 43\%, 44\%, and 57\% in the magnitude
range $24 < I_{814,AB} < 26$ and larger by 35\%, 42\%, and 53\% in the
magnitude range $26 < I_{814,AB} < 27.5$ for the $\Lambda =
0.1$/$\Omega = 0.1$, $\Omega = 0.1$, and $\Omega = 1$ geometries,
respectively, where the median of the angular size distribution in
each magnitude bin is used for comparison.  As expected, we see that
galaxies in the $\Lambda = 0.9$/ $\Omega = 0.1$ geometry tend to
possess smaller angular sizes than galaxies in the $\Omega = 0.1$
geometry and especially galaxies in the $\Omega = 1.0$ geometry
because of the somewhat larger angular-diameter distances.

We repeat the above comparisons using the Petrosian radii instead of
the half-light radii.  Ideally, Petrosian radii provide a more
reliable surface brightness-independent estimator of the angular size
than half-light radii, though one might question its meaningfulness
given the increasingly lumpy appearance and small sizes of faint
galaxies.  In any case, an inspection of Figure 15 shows that the same
general trends and conclusions hold here as for the half-light radii.

\begin{figure}
\epsscale{1.00}
\plotone{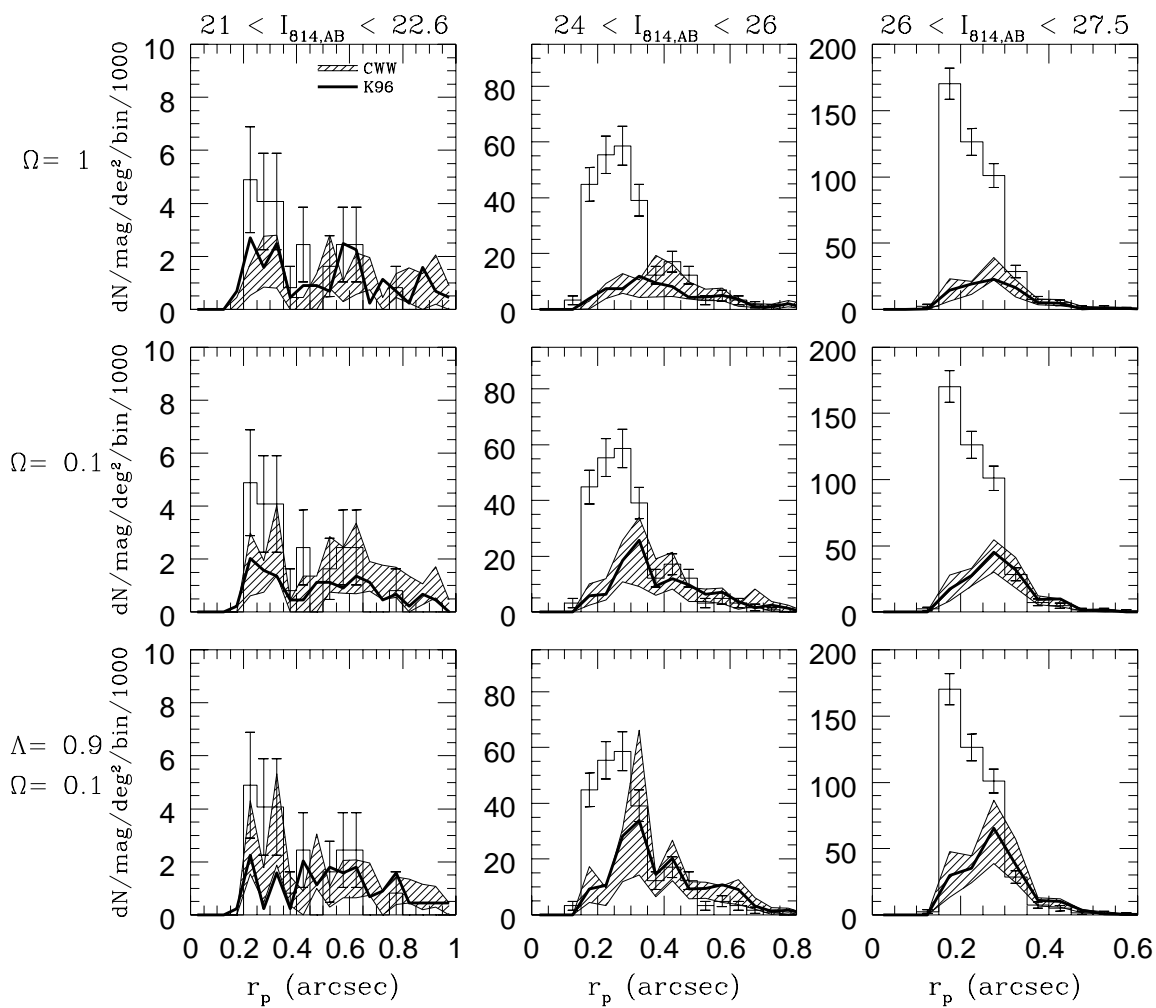}
\caption{Analogous to Figure 14 except using the Petrosian radius.
The observed distribution shows a sharper peak at small sizes, but the
basic conclusions are the same.  Note that contrary to the half-light
radii, the Petrosian radius is not sensitive to changes in the surface
brightness or the choice of isophotes.\label{angdistpa}}
\end{figure}

\placefigure{angdistpa}

\subsection{Colour Distributions}

In Figure 16-17, we plot a comparison between the colour distributions
recovered from the observations and our `no-evolution' simulations in
three different magnitude ranges for two different colours, $(B_{450}
- I_{814})_{AB}$ and $(V_{606} - I_{814})_{AB}$.  Good agreement is
found at the bright magnitudes ($21 < I_{814,AB} < 22.6$) as expected
given that our bright sample is selected up to this magnitude limit.
At fainter magnitudes, however, a clear excess of bluer galaxies is
observed relative to our no-evolution simulations.  In addition, in
the faintest magnitude bin ($26 < I_{814,AB} < 27.5$), there also
appears to be an excess of red galaxies relative to that found in the
no-evolution simulations.  This is somewhat unexpected because one
would expect the real universe, for which our sample is representative
at $z\sim0.5$ to have a younger and therefore bluer appearance than
that of our extrapolated sample.  This excess may, therefore, reflect
the presence of dust and Lyman-series forest absorption at moderate to
high-redshift.  Conceivably, an ad-hoc maximal dwarf-model, might also
account for this red excess in terms of faded dwarfs at low redshift.

\begin{figure}
\epsscale{0.90}
\plotone{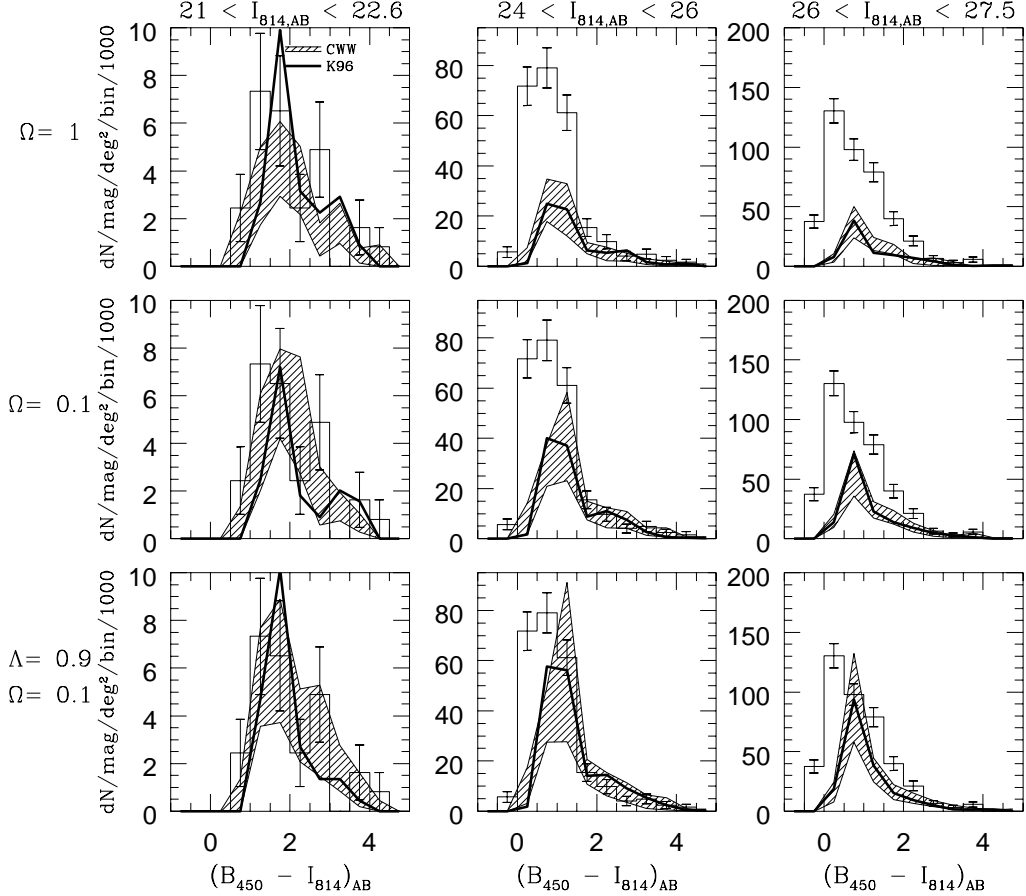}
\caption{Comparison of the colour distributions $(B_{450} -
I_{814})_{AB}$ for the observations (histogram with $1 \sigma$
Poissonian errors) with the simulations using the CWW SED templates
(hatched region representing the $1 \sigma$ uncertainties based upon
the finite size of our bright galaxy sample) and using the K96 SED
templates (solid line) for different geometries ($\Omega = 0.1$,
$\Lambda = 0.9$; $\Omega = 0.1$; $\Omega = 1$) and different magnitude
ranges.  Note the recovered distribution of colours from our
no-evolution simulations agrees roughly with the recovered
distribution of colours from the HDF in the bright magnitude bin as it
should since our no-evolution simulations are composed of precisely
these same galaxies.  At fainter magnitudes, the breadth of the
observed distribution is greater than the ``no-evolution''
simulations, with the mean shifted to bluer colours.
\label{colora}}
\end{figure}

\begin{figure}
\epsscale{1.00}
\plotone{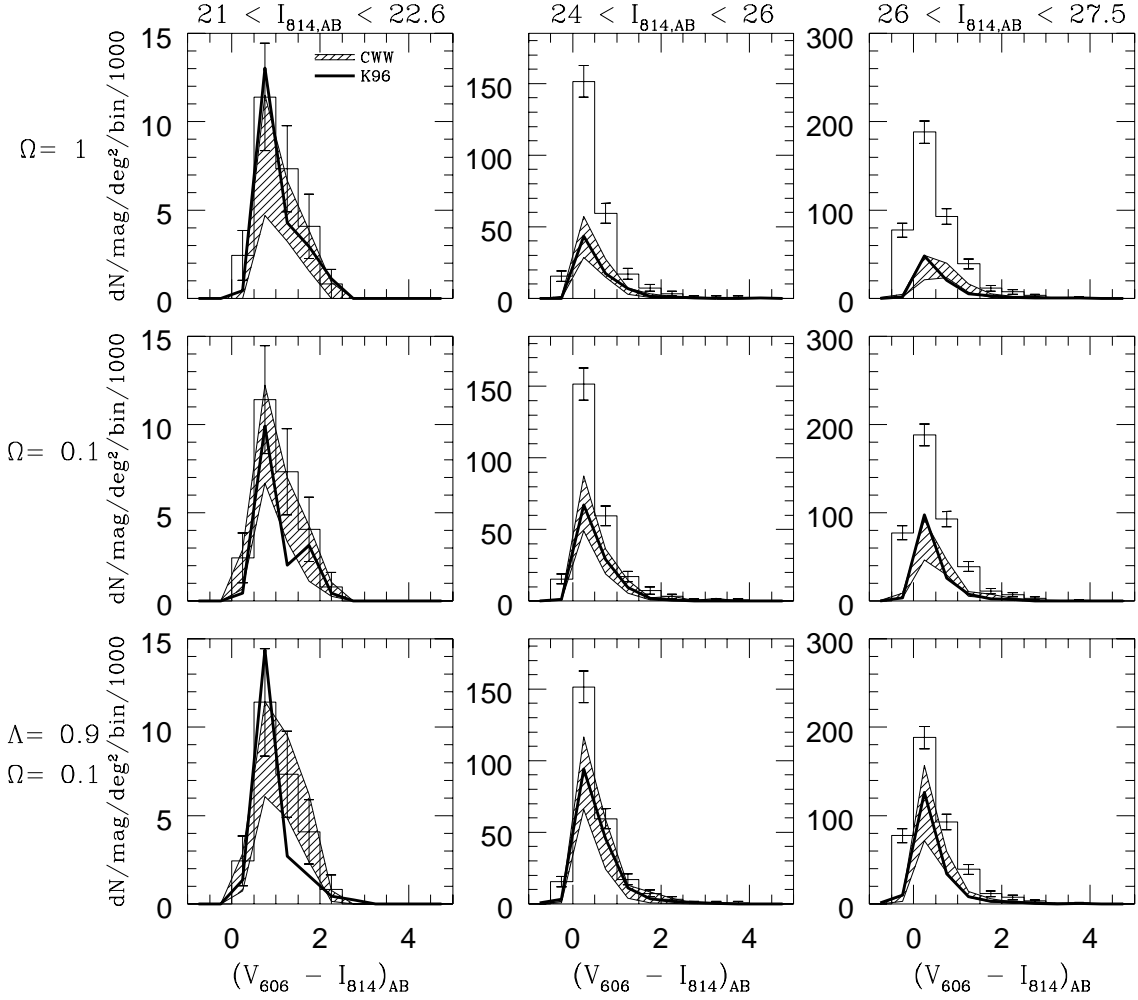}
\caption{Comparison of the colour distributions $(V_{606} -
I_{814})_{AB}$ for the observations (histogram with $1 \sigma$
uncertainties) with the simulations using the CWW SED templates
(hatched region representing the estimated 1 $\sigma$ uncertainties
based upon the finite size of the bright galaxy sample) and using the
K96 SED templates (solid line) for different geometries ($\Omega =
0.1$, $\Lambda = 0.9$; $\Omega=0.1$; $\Omega=1$) and different
magnitude ranges.
\label{colorva}}
\end{figure}

\placefigure{colora}
\placefigure{colorva}

\subsection{Redshift Distributions}

In Figure 18, we plot the redshift distributions of the galaxies
recovered by SExtractor by matching them up with our input catalogues.
It is apparent that in the absence of evolution very few galaxies lie
beyond a redshift $z=2$, even at the faintest magnitude, with little
dependence on the choice of geometry.  A dramatic illustration of the
relative unobservability of high redshift galaxies for $\Omega = 0.1$
is given in Figure 19, where we have broken up the simulation of a
52'' x 72'' HDF exposure into 4 redshift slices.

\begin{figure}
\epsscale{1.00}
\plotone{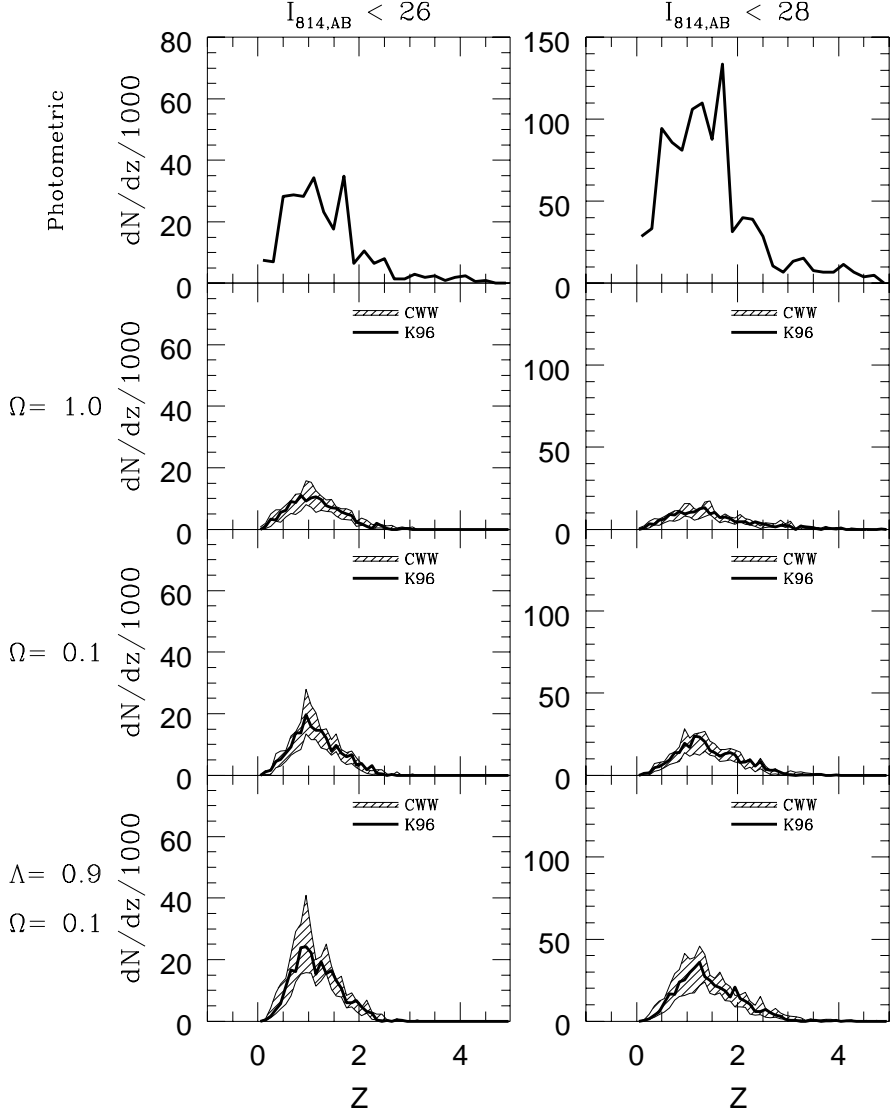}
\caption{Redshift distribution of those objects recovered by
SExtractor from no-evolution simulations with CWW SED templates
(hatched region indicating $1 \sigma$ uncertainties) and our
no-evolution simulations with K96 SED templates (solid line) with
$I_{814,AB} < 26$ and $I_{814,AB} < 28$ for $\Omega = 0.1$/$\Lambda =
0.9$, $\Omega = 0.1$ and $\Omega = 1$ geometries.  For comparison, the
upper panel shows the redshift estimates by Lanzetta, Yahil, \&
Fernandez-Soto (1996) which are lower than the estimates by Mobasher
et al.\ (1996) and higher than those of Sawicki, Lin, \& Yee
(1997).\label{obsz}}
\end{figure}

\begin{figure}
\begin{center}
\resizebox{11cm}{!}{\includegraphics*[100,150][450,650]{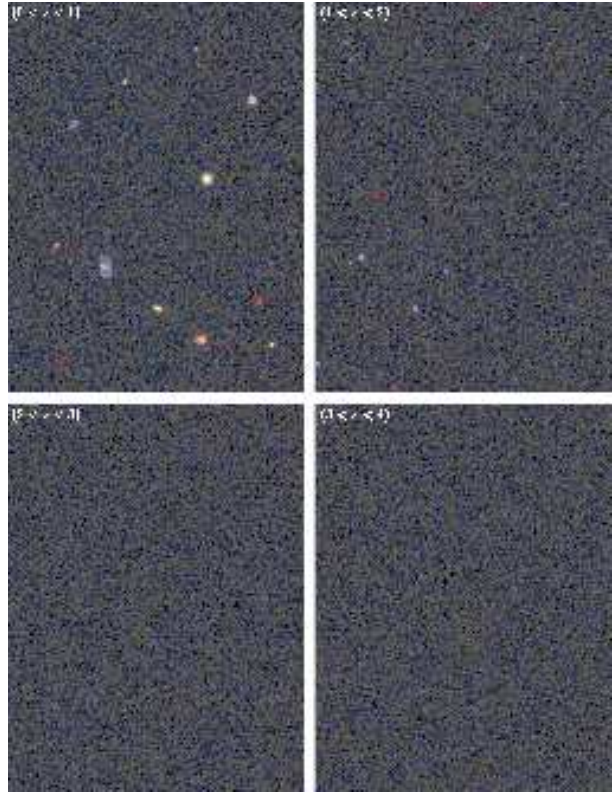}}
\end{center}
\caption{ Simulated 52'' x 72'' colour images generated from the
$B_{450}$, $V_{606}$, and $I_{814}$ bands for four different redshift
slices ($0 < z < 1$, $1 < z < 2$, $2 < z < 3$, and $3 < z < 4$) using
a $\Omega=0.1$ no-evolution simulation with pixel size,
signal-to-noise, and PSF identical to that of the HDF.  Note the
decreased visibility of our bright galaxy sample at high redshift.
\label{z4}}
\end{figure}

\placefigure{obsz}
\placefigure{z4}

It is interesting to see how these distributions compare with the
photometric redshift estimates derived from fits to the 4 passbands of
the HDF, and from subsequent near-IR imaging of this field from the
ground.  Surprisingly enough, the redshift distribution recovered from
our ``no-evolution'' simulations agree remarkably well with the
approximate distribution of Lanzetta, Yahil, \& Fernandez-Soto (1996).
Given that similar coincidences have been found for redshift surveys
at somewhat brighter magnitude limits (Broadhurst et al.\ 1988;
Colless et al.\ 1993; Cowie et al.\ 1996), it is somewhat tempting to
suppose that this result might hold to yet fainter magnitudes.
Despite such hopes, clearly this relationship begins to break down in
the $B$ band at fainter magnitudes ($B > 24$), at least, as Cowie et
al.\ (1996) have shown, and presumably a similar break down will occur
in the redder bands at fainter magnitudes.

\subsection{$U_{300}$ and $B_{450}$ ``Dropouts''}

In Figures 20 and 21, we plot several colour-colour diagrams,
and overplot the $U_{300}$-dropout and
$B_{450}$-dropout selection criteria given by Madau et al.\ (1996),
criteria, useful for selecting galaxies at redshifts $2 < z < 3.5$ and
$3.5 < z < 4.5$, respectively.  We count the number of recovered
galaxies satisfying these criteria and provide a sumary in 
Table 2.  As expected from the recovered redshift distributions
plotted in Figure 18, it is clear that the no-evolution simulations
contain manifestly fewer dropouts than the observations.  Clearly
then, the high-redshift universe contains a larger number of objects
which of greater luminosity in the ultraviolet than the galaxy population in
our redshift-complete sample.

\begin{figure}
\epsscale{0.80}
\plotone{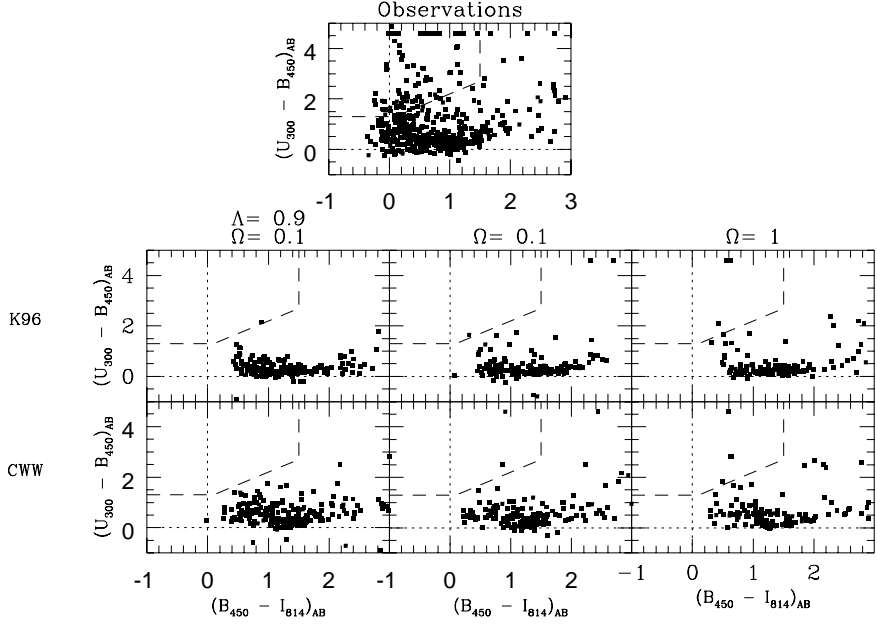}
\caption{Comparison of the $(U_{300} - B_{450})_{AB}$ versus $(B_{450}
- I_{814})_{AB}$ diagrams for our no-evolution simulations with the
CWW SED templates (lower panel) and our no-evolution simulations with
the K96 SED templates (top panel) with the observations for
$B_{450,AB} < 26.79$ (the same criterion used in Madau et al.\ 1996).
The area interior to the dashed line is the region Madau et al.\
(1996) suggests is occupied by high-redshift galaxies ($2<z<3.5$) whose
Lyman limit crosses the $U_{300}$ bandpass.  Note that galaxies near
the top of the colour-colour diagram, i.e., with $(U_{300} -
B_{450})_{AB} > 4.6$, are simply lower limits on the $(U_{300} -
B_{450})_{AB}$ colour.  These no-evolution simulations underpredict the
number of dropouts in this region.
\label{scu}}
\end{figure}

\begin{figure}
\epsscale{0.8}
\plotone{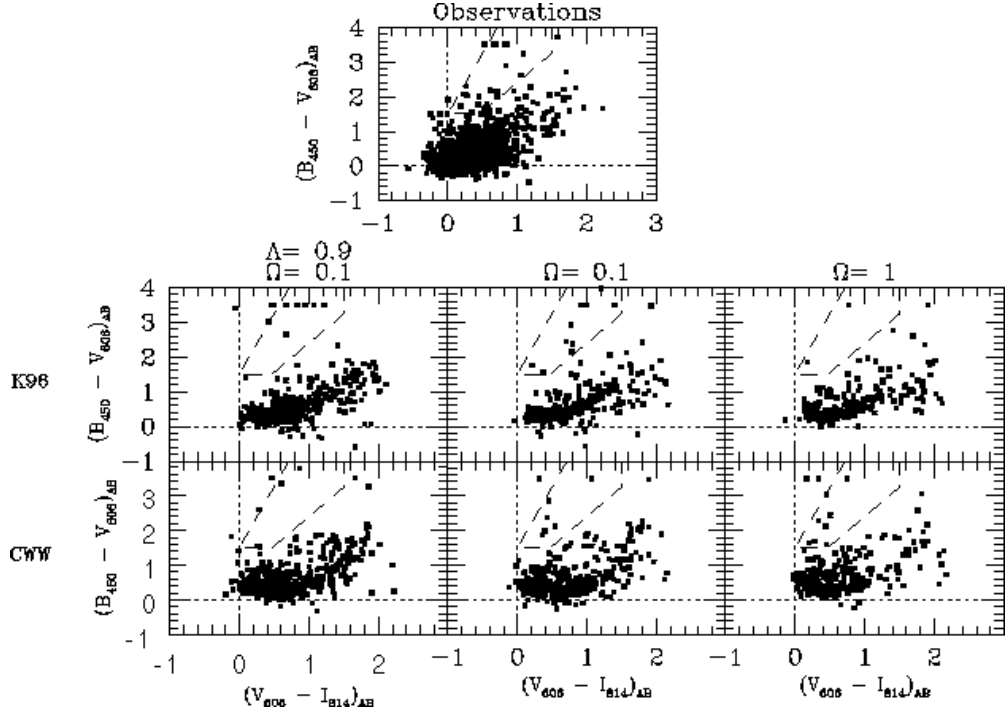}
\caption{Comparison of the $(B_{450} - V_{606})_{AB}$ versus
$(V_{606} - I_{814})_{AB}$ diagrams for our no-evolution simulations
with the CWW SED templates (lowest panel) and our no-evolution
simulations with K96 SED templates (top panel) against the
observations for galaxies with $V_{606,AB} < 28.0$ (the
same criterion used in Madau et al.\ 1996).  With a dashed line, we
have overplotted the $B$-band dropout region suggested by Madau et
al.\ (1996) for finding high redshift ($3.5 < z < 4.5$) galaxies whose
Lyman-limit crosses the $B$ bandpass.  Note that galaxies near the top
of the colour-colour diagram, i.e., with $(B_{450} - V_{606})_{AB} > 3.5$,
are simply lower limits on the $(B_{450} - V_{606})_{AB}$ colour.  As with
the $U_{300}$ dropouts, these no-evolution simulations underpredict
the number of dropouts.\label{sc}}
\end{figure}

\placefigure{scu}
\placefigure{sc}
\placetable{dropouts}

We have compared the numbers of dropouts in the HDF found by Madau et
al.\ (1996) with our values listed in Table 2, using the same
colour-magnitude window.  Our dropout rate is higher for the HDF, a 
finding which
might result from our use of SExtractor for photometry rather than
FOCAS as employed by Madau et al.\ (1996).  Given that these programs
likely differ with regard to the degree to which they successfully
estimate the magnitudes of detected objects, our analyses probably
probe slightly different depths, even though we choose for comparison
the same nominal magnitude limit ($I_{814} < 26.79$).  Imposing a
magnitude limit $0.^m3$ brighter, we can reproduce their dropout rate.
Also, as stated by Madau et al.\ (1996), their FOCAS magnitudes are in
general estimated to be too faint by about $0.^m5$.

\subsection{Asymmetry}

In the light of recent work attempting to quantify the morphology of
faint galaxies, we compare the properties of the faint galaxies from
the HDF against those from our simulations with the $A$ statistic
proposed by Abraham et al.\ (1996a,1996b), in order to provide an
approximate estimate of the extent to which evolution may affect the
asymmetry of galaxies, where $A$ is defined as the sum of the absolute
value of the differences between a galaxy image and itself rotated 180
degrees about the center of the image.  The approximate contribution
of the noise to the apparent asymmetry is calculated and subtracted.
We plot the distribution of this asymmetry parameter recovered both
from our simulations and from the HDF in Figure 22, for two different
magnitude bins.  We find a clear trend toward larger values of this
parameter with increasing magnitude.  At the simplest level, this
would seem to imply that the faint galaxies are less smooth than the
redshifted field population at $z\sim 0.5$ used in our simulations.

\begin{figure}
\epsscale{0.95}
\plotone{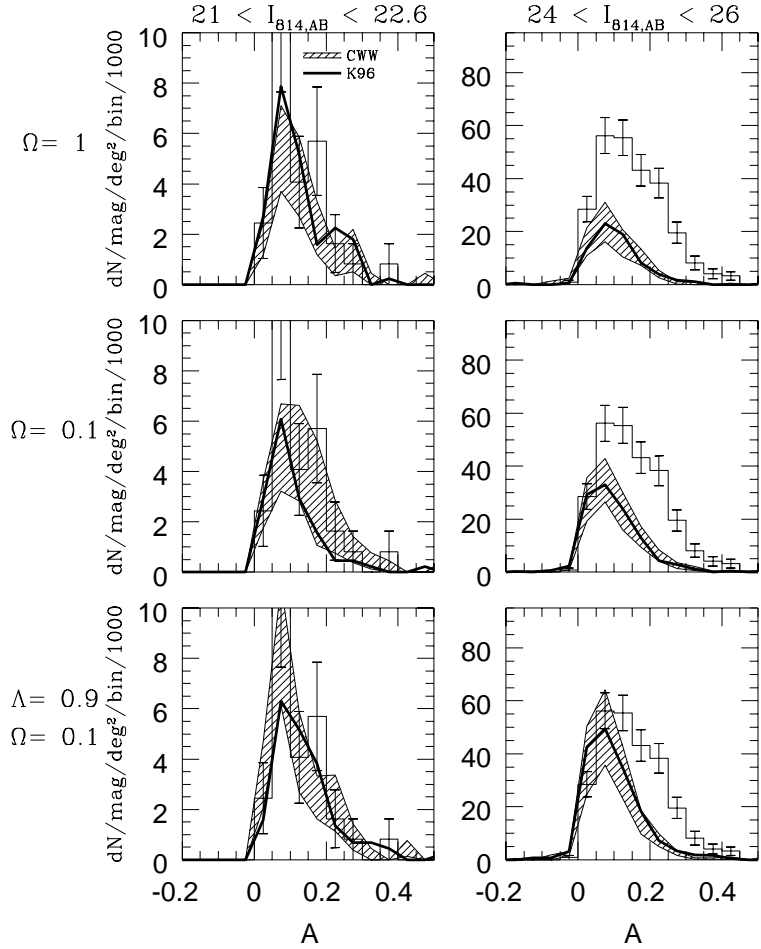}
\caption{Comparison of the distribution of asymmetries recovered from
the HDF (histogram with $1 \sigma$ Poissonian errors) and our
simulations assuming no-evolution using the CWW SED templates (hatched
region representing the 1 $\sigma$ uncertainties estimated from the
finite size of our bright sample) and using the K96 SED templates
(solid line).  The asymmetry statistic, pioneered by Abraham et al.\
(1996a,1996b), is equal to zero for completely symmetric objects and
systematically increases for more asymmetric objects.  Clearly, the
data appears to be systematically more asymmetric at fainter
magnitudes than the observations, though this result may be partially
biased by the general differences in angular-sizes of the populations
in question.
\label{adista}}
\end{figure}

\placefigure{adista}

Some caution must be exercised in interpreting the $A$ statistic since
it appears to be extremely sensitive to the manner in which one
determines the center about which to rotate for evaluating the
statistic (Abraham \& Brinchmann, private communication) and can
exhibit a fair amount of scatter ($\pm0.1$) depending upon whether one
takes the centroid or the maximum as the for the center.  Because of
this, the statistic, or at least the present implementation, seems to
become increasingly unreliable for the smallest galaxies and hence we
restrict our comparison to the bright magnitudes ($I_{814,AB}<26$)
where we are more confident in the reliability of the statistic from
our own internal tests.

Note that along similar lines, based upon pixel-by-pixel k-corrections
of the Frei, Guhathakurta, \& Gunn (1996) sample and systematic
applications of asymmetry ($A$) \& central-concentration ($C$)
statistics, Abraham et al.\ (1996a,1996b) have argued that the fainter
galaxy population is quite unlike the local galaxy population broadly
represented by the Frei et al.\ (1996) sample, which is nevertheless
somewhat ill-defined, lacking a firm magnitude limit or well-defined
selection criteria.

\section{Discussion}

Since the novelty of the HDF is principally its angular size
information obtained to unprecedentedly faint magnitudes, it is not
too surprising that our most interesting finding relates to the sizes
of the faint images.  The count excess is clearly composed of galaxies
with smaller projected areas than expected on the basis of relatively
low-redshift galaxies ($\bar{z}=0.5$) for any interesting geometry.
The evolution of angular-size has not been completely clear in the
literature to date.  Previously, in the range $24 < I < 25$, Roche et
al.\ (1996) find an apparent excess of galaxies with small angular
sizes relative to their no-evolution models which are designed to
replicate the angular sizes of galaxies in the local universe.  At
brighter magnitudes ($I \approx 22$), the study by Im et al.\ (1995)
on the prerefurbished HST data indicates that the angular sizes of
this population is also smaller than expected based upon more local,
brighter measures.  However, there have also been some studies based
on this parallel data indicating no changes in size (Mutz et al. 1994;
Casertano et al.\ 1995).

We also find that the number density of dropouts discovered at faint
magnitudes exceeds that recovered from our no-evolution simulation, a
finding which clearly indicates that the real high-redshift universe
contains a larger density of galaxies with high UV surface
brightnesses.  Note, however, that despite this finding, the estimated
integrated star-formation rates in these redshift ranges are similar
(Madau et al.\ 1996).  Clearly, then, as argued in Steidel et al.\
(1996) and Lowenthal et al.\ (1997) (cf. Figure 7), more
star-formation took place on relatively small scales at high-redshift.
It remains unclear whether this is the result of the intrinsically
smaller baryonic aggregations or the increasing prominence of
starbursts at high redshifts.

Our finding that overcounting rates do not seem to be very important
(see also the companion paper: Bouwens, Broadhurst, \& Silk 1998) is
somewhat contrary to the claims of Colley et al.\ (1997), even though
we used a similar program (SExtractor) with similar values of the
deblending parameter as one catalogue (Clements \& Couch 1997) used in
their study.  Of course, the galaxy population we have used for our
study is more evolved than expected in the real universe at higher
redshifts, so we might expect our value to be an underestimate. Note
that in any case, the small scale correlation claimed is a
statistically marginal result, ($\sim1.5\sigma$, Colley et al 1997).

Less contentious are our findings regarding the colours and the number
counts.  In agreement with most authors we find clear evidence that
the predictions without evolution are redder than observed (Lilly et
al.\ 1995) and that the number counts fall short of the data at faint
magnitudes (Lilly 1993; Pozzetti et al.\ 1996).

As we have already argued, it is possible to see the present work as a
demonstration that the trends found in comparing the local ($z\sim0$)
population with intermediate-redshift galaxies ($z\sim0.5$) extend to
yet fainter magnitudes and presumably higher redshifts.  At these
intermediate magnitudes a larger number of faint, blue, irregular, and
presumably small galaxies (Ellis et al.\ 1996; Lilly et al.\ 1995;
Brinchmann et al.\ 1998; Guzman et al.\ 1997) have been reported at
$z\sim0.5-0.8$ relative to local expectations.  The present study
demonstrates that the these trends extend to yet fainter magnitudes,
in that there continues to be both an increase in the number and
apparent starburst activity of the faint galaxy population as well as
a decrease in mean galaxy size.

Uncertainty estimates for the present study are somewhat ``internal''
and, therefore, might underestimate a field-averaged variance in our
measurements.  Certainly, at bright magnitudes the HDF is $\sim28\%$
overdense with respect to the CFRS.  However, since 90\% of our bright
galaxy sample ranges over a wide spread in distance, some 1500 Mpc/h,
and $V/V_{max}$ is evenly spread over the interval [0,1], most bright
galaxies should be considered as independent.  Clearly, we would
prefer a larger, more local sample of redshift-complete bright
galaxies for our simulations, and other deep multicolour HDF imaging
would be useful in the first two respects.

\section{Summary}

We have developed a technique for generating a model-independent faint
galaxy population based on pixel-by-pixel k-corrected images of the
bright galaxy population, for comparison with deep high-resolution
images. Our technique has the virtue of being model-independent and
completely empirical, except for the usual choice of geometry.  We
have made use of the HDF for the purposes of defining a
redshift-complete bright galaxy sample from which to construct
empirical simulations and also for evaluating the evolution of the
much fainter galaxies detected in the HDF.  We have made use of all
four passbands and have been able to make concrete statements about
the evolution of the image properties with estimated uncertainties
based on the size of a volume-limited sample.  We find that, relative
to our simulations based on the bright galaxy sample, the faint
galaxies are smaller, more numerous, bluer, less regular, and contain
more dropouts for any interesting geometry.  Of particular note is
our finding with regard to angular sizes, for which a variety of
nebulous and seemingly contradictory statements may be found in the
literature.

Our bright galaxy sample is small, and it will be important to see if
our conclusions remain valid when a larger sample is available.
Nonetheless, we believe this work is an important foundation on which
to build, in particular, as we look forward to independent deep fields
and the very significant improvement in the quality of UV and optical
imaging promised by the Advanced Camera (Ford et al.\ 1996).

\acknowledgements

We would like to thank Daniela Calzetti, Marc Davis, Mike Fall,
Holland Ford, Andy Fruchter, Nick Kaiser, and Alex Szalay for some
very useful conversations, Emmanuel Bertin for answering several of
our questions regarding SExtractor, Gordon Squires and Nick Kaiser for
the use of several routines from their software package IMCAT, Harry
Ferguson for his help in producing colour images and answering a few
of our questions, Jarle Brinchmann and Roberto Abraham for a number of
useful tips and comments with regard to the routine we incorporated
into SExtractor for measuring the $A$ parameter, and finally Harry
Ferguson, Steve Zepf, Eric Gawiser, and Jonathan Tan for some helpful
comments on near-final drafts of this document.  RJB acknowledges
support from an NSF graduate fellowship, TJB acknowledges the NASA
grant GO-05993.01-94A, and JS acknowledges support from NSF and NASA
grants.

\appendix

\section{Determination of Random Errors}

Here we quantify the empirical model uncertainties due to the finite
size of our input sample.  We look at this uncertainty in terms of an
arbitrary quantity describing the surface density of objects on the
sky satisfying some observational criteria.

For any faint sample of objects satisfying a certain set of observable
criteria ($S_O$), it is possible to express the probability
distribution for the number of these objects as the sum of the
probability distributions for this variable over all possible galaxy
morphologies, luminosities, and sizes, i.e., $\sum_{i=1} ^N f_i$.  We
suppose that this faint sample of galaxies is derived in a Monte-Carlo
manner from our bright sample on the basis of no-evolution
assumptions, so that it is possible to express this probability
distribution for each galaxy type as the product of the probability
distribution for the number of times this object will appear in our
bright sample, call it $B_i$, and the probability distribution for the
number of times such a galaxy would be expected to appear in the faint
sample in question ($S_O$), call it $D_i$.  In this way, the overall
probability distribution for the number of galaxies observed in a
given faint sample ($S_O$) can be expressed as
\begin{displaymath}
\sum_{i=1} ^N B_i D_i
\end{displaymath}
We shall suppose that each galaxy in our bright sample is sufficiently
unique to be represented by its own term in the above equation.
Furthermore, since each of these galaxies appeared in our bright
sample once and only once, we suppose that $B_i$ is
Poisson-distributed with a mean of 1, which we approximate
as a normal distribution about this mean.  For lack of something
better, we shall take the probability distributions $B_i$ for all
other galaxy types to be delta functions at 0.\footnote{ Note however
that galaxy prototypes with relatively large values of $D_i$, i.e.,
with relatively larger contributions at faint magnitudes than at
bright magnitudes, could result in large systematic errors in this
distribution.}

Now, it is possible to estimate the uncertainty in the expectation
value for the number of galaxies in our faint sample based on the
finite number of galaxies we have in our bright sample.  Because our
simulated area is, in principle, unlimited, the uncertainty in the
expectation value of $D_i$ is vanishingly small, and in practice we
take this uncertainty to be zero.  As such, the above
probability distribution is simply the sum of normal distributions
with various weights, and therefore its 1-sigma uncertainty is
simply equal to $\sqrt{\sum_{i=1} ^N D_i^2}$.  We have used this
expression throughout our paper in the estimation of errors for each
observable calculated from our no-evolution simulations.

\section{Breakdown of Faint Samples}

At fainter magnitudes, one expects a smaller number of the galaxy
prototypes to make up an increasing fraction of our faint galaxy
sample, as a result both of the differential k-corrections and of the
relatively greater volume available to lower luminosity objects at
fainter magnitudes.  To quantify the importance of this trend, we
compute a weighted measure of the effective number of galaxies which
contribute to the counts in each magnitude bin, a quantity we take to
equal
\begin{displaymath}
\left(\frac{\sum_{i=1} ^N n_i}{\sqrt{\sum_{i=1} ^N n_i^2}}\right)^2
\end{displaymath}
where $n_i$ is the number of galaxies of a given prototype per
magnitude per square degree recovered from our simulations and $N$ is
the number of prototypes considered (i.e. 31).  Basically, the above
expression equals the square of the number counts divided by the
estimated Poissonian error based on the size of the contributing
sample at the magnitude in question, the derivation of which is given
in Appendix A.  In the case that all the $n_i$'s are the same, the
expression reduces to $N$ as required.

We compute the above expression from our simulations, and we plot
the results in Figure 23.  At bright magnitudes, the bright galaxy
catalogues are still dominated by shot noise so that the number of
galaxies estimated from the above expression to contribute at any
magnitude is lower than the real number ($\sim 31$) which contribute.
At fainter magnitudes, the effective of number of galaxies decreases
again for the reasons stated above, i.e., a relatively smaller
differential volume for the more luminous galaxies and the
k-correction.  For the sake of clarity, we emphasize that at each
magnitude, every galaxy prototype contributes some fraction to the
number counts there.

\begin{figure}
\epsscale{1.0}
\plotone{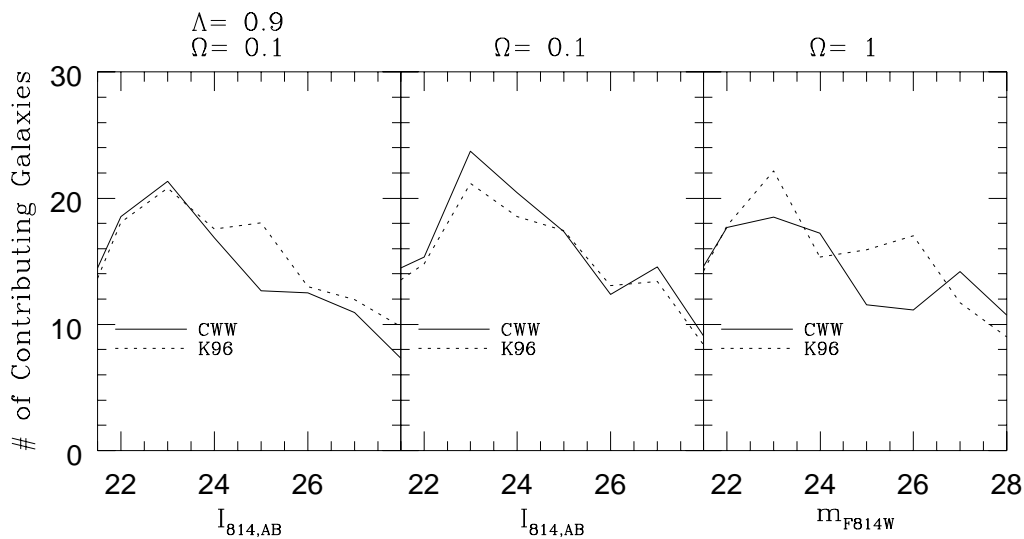}
\caption{This figure sows the effective number of the galaxy
prototypes which contribute as a function of magnitude (see Appendix B
for a description).  At brighter magnitudes, the effective number
should be close to the number of galaxy prototypes, the shortfall seen
here can be attributed to shot noise.  At fainter magnitudes, because
of a relatively smaller differential volume for the more luminous
galaxies and differing k-corrections, the bluer galaxy prototypes
contribute more to the counts than others, effectively reducing the
numbers of prototypes sampled at higher redshift.
\label{ncont}}
\end{figure}

\placefigure{ncont}

{}

\newpage

\newpage

\newpage

\newpage

\newpage

\newpage

\clearpage

\newpage

\newpage

\newpage

\newpage

\newpage

\newpage

\newpage

\newpage

\newpage

\newpage

\newpage

\newpage

\newpage

\newpage

\newpage

\newpage

\clearpage
\newpage

\begin{deluxetable}{lrrrrrrr}
\tablewidth{0pt}
\tablecaption{Sample of HDF Galaxies Used.\label{tbl-1}}
\tablehead{
\colhead{Right Ascension} & \colhead{Declination} &
\colhead{z} &
\colhead{$I_{814,AB}$} & \colhead{$M_{b_j}$\tablenotemark{a}} 
& \colhead{$k(2.5)$\tablenotemark{b}}
& \colhead{$\mu$ ($b_j$)\tablenotemark{c}} & \colhead{MT}\tablenotemark{d}}

\startdata
12:36:49.351 & 62:13:47.934 & 0.089 & 18.2 & -19.2 &  4.4 & 19.7 & E\\
12:36:50.971 & 62:13:21.738 & 0.199 & 19.6 & -20.2 &  1.0 & 21.9 & Sp\\
12:36:56.572 & 62:12:46.478 & 0.517 & 20.1 & -21.8 &  4.1 & 21.6 & Sp\\
12:36:47.992 & 62:13:10.107 & 0.475 & 20.5 & -21.2 &  4.3 & 19.9 & E\\
12:36:50.138 & 62:12:40.811 & 0.474 & 20.7 & -21.4 &  1.1 & 20.7 & Mrg\\
12:36:42.827 & 62:12:17.400 & 0.454 & 20.7 & -21.1 &  1.2 & 20.5 & Sp\\
12:36:41.654 & 62:11:32.961 & 0.089 & 20.8 & -18.4 &  0.6 & 21.9 & Sp\\
12:36:43.713 & 62:11:43.907 & 0.764 & 20.9 & -22.5 &  3.8 & 19.2 & E\\
12:36:53.820 & 62:12:55.070 & 0.642 & 20.9 & -21.8 &  2.0 & 20.4 & Sp\\
12:36:41.850 & 62:12:06.478 & 0.432 & 21.0 & -20.7 &  1.5 & 20.6 & Sp\\
12:36:46.952 & 62:12:37.901 & 0.318 & 21.0 & -19.9 &  0.9 & 20.7 & E\\
12:36:51.708 & 62:13:54.832 & 0.557 & 21.1 & -21.2 &  2.1 & 20.5 & Ir\\
12:36:46.082 & 62:11:43.119 & 1.012 & 21.3 & -23.1 &  1.5 & 20.3 & Sp\\
12:36:58.690 & 62:12:53.457 & 0.319 & 21.3 & -19.7 &  1.0 & 22.6 & Sp\\
12:36:46.262 & 62:14:05.706 & 0.960 & 21.3 & -23.0 &  2.9 & 18.5 & E\\
12:36:57.230 & 62:13:00.701 & 0.473 & 21.4 & -20.8 &  0.5 & 21.7 & Sp\\
12:37:00.485 & 62:12:35.720 & 0.562 & 21.4 & -20.8 &  3.3 & 20.1 & E\\
12:36:50.193 & 62:12:46.807 & 0.677 & 21.5 & -21.5 &  4.4 & 20.1 & E\\
12:36:49.634 & 62:13:14.095 & 0.475 & 21.5 & -20.4 &  2.3 & 21.7 & Sp\\
12:36:44.287 & 62:11:34.292 & 1.013 & 21.6 & -23.5 &  4.7 & 18.9 & E
\tablebreak
12:36:51.640 & 62:12:21.247 & 0.299 & 21.6 & -19.0 &  2.5 & 21.1 & Sp\\
12:36:44.097 & 62:12:48.868 & 0.555 & 21.7 & -20.8 &  0.4 & 20.6 & Sp\\
12:36:43.066 & 62:12:43.258 & 0.847 & 21.8 & -22.2 &  4.8 & 18.9 & E\\
12:36:49.416 & 62:14:07.778 & 0.752 & 21.8 & -21.4 &  1.7 & 19.5 & E\\
12:36:55.493 & 62:12:46.551 & 0.790 & 22.0 & -21.5 &  1.6 & 20.4 & Sp\\
12:36:46.422 & 62:11:52.360 & 0.504 & 22.0 & -19.9 &  3.5 & 19.6 & E\\
12:36:49.559 & 62:12:58.629 & 0.475 & 22.0 & -19.9 &  1.7 & 20.7 & Sp\\
12:36:49.288 & 62:13:12.300 & 0.478 & 22.2 & -19.9 &  0.9 & 19.3 & E\\
12:36:43.936 & 62:12:50.496 & 0.556 & 22.2 & -21.3 &  2.0 & 20.5 & Irr\\
12:36:38.882 & 62:12:20.817 & 0.608 & 22.2 & -20.4 &  1.0 & 21.2 & Sp\\
12:36:39.920 & 62:12:08.424 & 1.015 & 22.3 & -22.5 &  4.5 & 18.3 & E\\
\enddata
\tablenotetext{a}{Assuming $\Omega = 0.1$, $H_0$ = 50 km/s/Mpc, 
and CWW SEDs (A0V magnitudes)}
\tablenotetext{b}{K-correction in $I_{814}$ at redshift 2.5 using CWW SEDs}
\tablenotetext{c}{Central surface brightness taken to equal $m_{b_J}
 (z=0)/(2 \pi r_{hl} ^2)$ (A0V magnitudes)}
\tablenotetext{d}{Our own eyeball classification of the morphological
type}
\end{deluxetable}

\begin{deluxetable}{ccc}
\tablewidth{0pt}
\tablecaption{Number of $U_{300}$ and $B_{450}$ dropouts.  One $\sigma$
uncertainties are given on all simulated results based on the finite
size of our bright sample.\label{dropouts}}
\tablehead{ \colhead{Data set} & \colhead{$U_{300}$ dropouts} & 
\colhead{${B_{450}}$ dropouts}}
\startdata
Observations (Madau et al.\ 1996) & 58 & 14 \\
Observations (This work) & 90 & 19 \\
NE ($\Omega = 0.1$/$\Lambda = 0.9$/CWW) & $ 1\pm 1$ & $3 \pm 2$ \\
NE ($\Omega = 0.1$/CWW) & $2 \pm 1$ & $4 \pm 3$ \\
NE ($\Omega = 1$/CWW) & $6 \pm 3$ & $5 \pm 3$ \\
NE ($\Omega = 0.1$/$\Lambda = 0.9$/K96) & $2 \pm 2$ & $3 \pm 2$ \\
NE ($\Omega = 0.1$/K96) & $2 \pm 1$ & $4 \pm 3$ \\
NE ($\Omega = 1$/K96) & $3 \pm 2$ & $2 \pm 2$ \\
\enddata
\end{deluxetable}

\end{document}